\documentclass[fleqn,usenatbib]{mnras}  % a4paper
\usepackage{newtxtext,newtxmath}
\usepackage{xfrac}

\usepackage{ae,aecompl}

\usepackage{graphicx}	% Including figure files
\usepackage{amsmath}	% Advanced maths commands
\usepackage[dvipsnames]{xcolor} % colors
\usepackage{ulem}
\usepackage{microtype}
\newcommand{\aref}[1]{\hyperref[#1]{Appendix~\ref{#1}}}

\newcommand{\zstart}{z_{\mathrm{start}}}
\newcommand{\astart}{a_{\mathrm{start}}}
\newcommand{\solarmass}{\mathrm{M}_{\odot}}
\newcommand{\dd}{{\rm d}}
\newcommand{\vect}[1]{\boldsymbol{#1}}

\newcommand{\ihMpc}{ h\,{\rm Mpc}^{-1}}

\usepackage{tikz}
\definecolor{lime}{HTML}{A6CE39}
\DeclareRobustCommand{\orcidicon}{
    \begin{tikzpicture}
    \draw[lime, fill=lime] (0,0) 
    circle [radius=0.14] 
    node[white] {{\fontfamily{qag}\selectfont \tiny ID}};
    \draw[white, fill=white] (-0.0625,0.095) 
    circle [radius=0.007];
    \end{tikzpicture}
    \hspace{-2mm}
}

\foreach \x in {A, ..., Z}{%
    \expandafter\xdef\csname orcid\x\endcsname{\noexpand\href{https://orcid.org/\csname orcidauthor\x\endcsname}{\noexpand\orcidicon}}
}

\title[Cosmic initial conditions: 3LPT]{Accurate initial conditions for cosmological $\boldsymbol{N}$-body simulations: \\ Minimizing truncation and discreteness errors}

\author[M.~Michaux et al.]{
Micha\"el Michaux$^{\,{\tiny\orcidA{}}\,\,\,\,\hyperlink{OCA}{1}}$\thanks{E-mail: michael.michaux@oca.eu}, Oliver Hahn$^{\,{\tiny\orcidB{}}\,\,\,\,\hyperlink{OCA}{1}}$\thanks{E-mail: oliver.hahn@oca.eu}, Cornelius Rampf$^{\,{\tiny\orcidC{}}\,\,\,\,\hyperlink{OCA}{1}}$ and Raul E. Angulo$^{\,{\tiny\orcidD{}}\,\,\,\,\hyperlink{OCA}{\,2,3}}$
\\
$^{1}$Universit\'e C\^ote d'Azur, Observatoire de la C\^ote d'Azur, CNRS, Laboratoire Lagrange, \\\quad Boulevard de l'Observatoire, CS 34229, 06304 Nice, France\\
$^{2}$Donostia International Physics Center (DIPC), Paseo Manuel de Lardizabal, 4, 20018 Donostia-San Sebasti\'an, Spain\\
$^{3}$IKERBASQUE, Basque Foundation for Science, 48013, Bilbao, Spain.}

\date{Accepted XXX. Received YYY; in original form ZZZ}

\pubyear{2020}

\begin{document}\label{firstpage}
\pagerange{\pageref{firstpage}--\pageref{lastpage}}
\maketitle

\begin{abstract}
	Inaccuracies in the initial conditions for cosmological \(N\)-body simulations could easily be the largest source of systematic error in predicting the non-linear large-scale structure. From the theory side, initial conditions are usually provided by using low-order truncations of the displacement field from Lagrangian perturbation theory, with the first and second-order approximations being the most common ones. Here we investigate the improvement brought by using initial conditions based on third-order Lagrangian perturbation theory (3LPT). We show that with 3LPT, truncation errors are vastly suppressed, thereby opening the portal to initializing simulations accurately as late as \(z=12\) (for the resolution we consider). We analyse the competing effects of perturbative truncation and particle discreteness on various summary statistics. Discreteness errors are essentially decaying modes and thus get strongly amplified for earlier initialization times. We show that late starting times with 3LPT provide the most accurate configuration, which we find to coincide with the continuum fluid limit within 1~per cent for the power- and bispectrum at \(z=0\) up to the particle Nyquist wave number of our simulations (\(k \sim 3h/\)Mpc). In conclusion, to suppress non-fluid artefacts, we recommend initializing simulations as late as possible with 3LPT. We make our 3LPT initial condition generator publicly available.
\end{abstract}


%%%%%%%%%%%%%%%%%%%%%%%%%%%%%%%%%%%%%%%%%%%%%%%%%%
%%%%%%%%%%%%%%%%%%%%%%%%%%%%%%%%%%%%%%%%%%%%%%%%%%

\section{Introduction}

Current and upcoming space and ground-based galaxy clustering and weak gravitational lensing surveys -- such as HSC \citep{hsc}, the LSST \citep{lsst}, and the Euclid satellite \citep{euclid}, but also future instruments probing the gas distribution across cosmic history, such as the Square Kilometre Array \citep{Bull:2018lat}, will test the validity of, and possible deviations from, the concordance \(\Lambda\)CDM model with unprecedented precision.

The accuracy of future observational data is expected to be such that it will require one-per-cent-accurate predictions for the spatial distribution of matter in the late-time Universe to scales of $k\sim1\,{\rm Mpc}^{-1}$ \cite[e.g.][]{Heitmann:2010,Schneider:2016}. Numerical simulations are, in principle, able solve the relevant equations with such accuracy over volumes comparable to the entire visible Universe \citep[e.g.][]{Angulo:2012,Heitmann:2016,Potter:2017,Heitmann:2019,Cheng:2020}.
However, numerical solutions are computationally demanding and suffer from numerical artefacts and discretization errors.
On the other hand, perturbative approaches of the underlying equations are not affected by such effects and deliver computationally cheap predictions.
Yet their validity is limited to certain length and time scales,
since the gravitational collapse of cosmic structure is intimately tied to the development of extreme densities that still pose a challenge for such approaches.

The worlds of analytic and numerical approaches come together when the formation of cosmic structures is investigated via cosmological simulations. Indeed, perturbative approaches essentially solve for the structure formation at early times, while numerical solutions are best at evolving later stages. Consequently,   perturbation theory is employed to generate perturbatively truncated initial conditions (ICs) for cosmological simulations \citep[as pioneered by][]{Klypin:1983,Efstathiou:1985}. Since both perturbation theory and numerical solutions have their strengths and weaknesses (see further below), it is important to find the optimal window where ICs can be provided while simultaneously minimizing perturbative and numerical errors. Quantifying these errors and finding this optimal cosmic time for ICs are precisely the tasks of this paper.

Observations indicate that cold dark matter (CDM) is extremely weakly (self-)interacting,  implying that it can be treated to be effectively collisionless with zero temperature on cosmological scales. The evolution of a continuous and collisionless medium, such as CDM, is governed by the Vlasov--Poisson equations. Most perturbative approaches solve these equations in the single-stream limit (vanishing velocity dispersion), which appears to be well justified for sufficiently early times. There exist four exact solutions in the single-stream limit, namely for one-dimensional ICs called the Zel'dovich approximation \citep[ZA; ][]{Zeldovich:1970}, for quasi-one dimensional ICs~\citep{Rampf:2017jan}, for spherical collapse~\citep{1967ApJ...147..859P}, and for quasi-spherical ICs~\citep{Rampf:2017tne}. Except for the spherical case, those solutions are valid until, and including, the instance of the first shell-crossing, where particle trajectories overlap, leading to the generation of (effective) vorticity and velocity dispersion through multi-streaming~\citep[e.g.][]{Pichon:1999,Pueblas:2009,Hahn:2014lca,Buehlmann:2019}.
Most recently, attempts at pushing the theory beyond shell-crossing have received increasing interest~\citep{Colombi:2014lda,Taruya:2017,McDonald:2018,Pietroni:2018,Rampf:2019nvl,Valageas:2020mzj}. For a broader overview over cosmological perturbation theory (PT), we refer the reader to the review by \cite{Bernardeau:2002}.

In general the non-linear evolution of the post-shell crossing regime is currently most accurately modelled with \(N\)-body simulations of the Vlasov--Poisson equations. The transition, via ``initial conditions'', from perturbation theory to the full non-linear but discretized simulation is, however, a rather delicate matter since it involves minimizing the errors in the approaches.

Initial conditions for simulations are usually provided by particle displacements and velocities from Lagrangian perturbation theory (LPT), truncated at some low order. This truncation, however, leads to so-called `transients' \citep{Scoccimarro:1998,Crocce:2006}; effectively this is a spurious decaying mode due to missing terms relative to the true (infinite-order) solution.
One alternative to ameliorate the impact of transients is to consider higher order versions of LPT. In fact, LPT has been derived at increasingly higher order over the last decades; starting from first-order by \cite{Zeldovich:1970,Buchert:1987xy}, over second \citep{1992ApJ...394L...5B,Buchert:1993xz} and third order \citep{Buchert:1994,Bouchet:1995} in the 1990s, to fourth order by \cite{Rampf:2012a,Tatekawa:2014}, and finally all-order recursion relations by~\cite{Rampf:2012b,Zheligovsky:2014,Matsubara:2015ipa}. Another alternative could be to generate initial conditions at earlier times since the linear growth amplitude of density fluctuations in LPT, $D_+$, is the small perturbative parameter.

Unfortunately, starting numerical simulations at early times might add significant sources of numerical error. As is now well-known, the \(N\)-body method is prone to discreteness effects due to the self-interaction of the discrete particle lattice \citep[cf.][]{Joyce:2005,Joyce:2007,Garrison:2016}, leading to a deviation from the fluid characteristics in the continuum limit. New tessellation methods reduce particle discreteness and thus could overcome these limitations \citep{Hahn:2013,HahnAngulo:2016,Sousbie:2016,Stuecker:2019}, albeit at increased computational cost. However, this might not be possible when simulations are optimized to simulate volumes as large as possible with the least possible computational resources. Recently, \cite{Garrison:2016} have argued for an explicit linear-order correction during the course of a simulation, but this is not widely used. Furthermore, popular methods such as tree-based \(N\)-body \citep{Barnes:1986} can suffer from large force errors at early times (where the density distribution is only slightly perturbed). In addition, these errors accumulate the more time steps are made during the early (``linear'') stages of a simulation when the density field is still close to homogeneous.

{\sl These problems would strongly suggest starting a simulation as late as possible. However, since standard perturbative approaches break down at shell-crossing, a competition exists between (i) late starts that reduce discreteness effects and numerical errors, but require higher-order LPT, and (ii) early starts that allow lower-order LPT but are more prone to discreteness errors. This dilemma is our main focus in this paper.}

The impact of the order of the LPT, up to second order, and the starting time on properties of the non-linear density field have been studied already in quite some detail in the past literature \citep[e.g.][]{Crocce:2006,Tatekawa:2007,LHuillier:2014,Garrison:2016}. Here, we extend these results to third-order LPT, which has been studied much less \citep[see however][]{Buchert:1993df,Tatekawa:2014,Tatekawa:2019}, thereby allowing us to make for the first time more definite statements about the convergence radius and thus the perturbative regime accessible by LPT for cosmological ICs. In our analysis we pay particular attention also to the impact of aforementioned discreteness effects on all results.

This paper is organized as follows. We begin with a brief review of LPT together with explicit solutions up to third order; see~\autoref{sec:LPT}. We implement these LPT solutions numerically, which allows us to perform simple convergence tests, thereby essentially pinning down until which time-value LPT solutions to any order can be trusted. Details and results on this are provided in~\autoref{sec:convergence_radius}.
Then,  in~\autoref{sec:numerical_results}, we discuss the details of our numerical simulations and analysis algorithms, whereas our results are given in~\autoref{sec:results}. We conclude and summarize our results in \autoref{sec:summary}.

Notation: Unless otherwise stated, all functions and spatial derivatives are w.r.t.\ Lagrangian coordinate $\vect{q}$. We use \(i\), \(j\), \ldots for spatial indices, and summation over repeated indices is assumed. A comma ``${\Phi}_{,i}$'' denotes a spatial partial derivative w.r.t.\ component $q_i$ on $\Phi$, while an overdot denotes a Lagrangian time derivative w.r.t.\ the cosmic-scale factor time $a$, the latter governed by the usual Friedmann equations.

%%%%%%%%%%%%%%%%%%%%%%%%%%%%%%%%%%%%%%%%%%%%%%%%%%
\section{LPT, initial conditions and its numerical implementation}\label{sec:LPT}

In this section we will discuss several theoretical and practical aspects of the creation of initial conditions for cosmological simulations.
Specifically, in \S\ref{sec:LPT_intro} we begin with a review of LPT solutions up to third order, and describe in  \S\ref{sec:backscaling} how these solutions are employed to create a consistent particle representation at a given starting time. We discuss the technical aspects of higher-order LPT implementations in  \S\ref{sec:numerical_implementation_lpt}. The effects of particle discreteness and initial arrangement are outlined and validated in \S\ref{sec:lattice_PLT} and \S\ref{sec:validation}, respectively.

\subsection{LPT results to third order}\label{sec:LPT_intro}

Let $\vect{q} \mapsto \vect{x}(\vect{q}, t) = \vect{q} + \vect{\psi}(\vect{q},t)$ be the Lagrangian map from initial position $\vect{q}$ to current (Eulerian) position $\vect{x}$ at time $t$. The Lagrangian representation of the velocity is defined with $\vect{v}\equiv \dot{\vect{x}}=\dot{\vect{\psi}}$.
In LPT, the displacement field, $\vect{\psi}$, is expanded as a power series in $D_+ = D_+(t)$, the linear growth of matter fluctuations in a \(\Lambda\)CDM universe, i.e.,
\begin{equation}\label{ansatz}
	\vect{\psi}(\vect{q},t) = \sum_{n=1}^\infty \vect{\psi}^{(n)}(\vect{q})\,D_+^n \,.
\end{equation}
The truncation of the series at order $n$ is commonly called $n$LPT, except for the first-order truncation which is called the Zel'dovich approximation \citep[ZA;][]{Zeldovich:1970}. Results to third order have been first derived by~\cite{Buchert:1994,Catelan:1995,Bouchet:1995}.
Explicitly, the 3LPT solution for the displacement is
\begin{equation}\label{eq:displacement_field}
	\vect{\psi}_{\rm 3LPT}(\vect{q},t) = \vect{\psi}^{(1)}(\vect{q})\, D_+ + \vect{\psi}^{(2)}(\vect{q})\, D_+^2 + \vect{\psi}^{(3)}(\vect{q})\, D_+^3 \,,
\end{equation}
with
\begin{align}\label{eq:3lpt_solutions}
	\vect{\psi}^{(1)} & =  -\nabla \Phi^{(1)} \,,                                                                                        \\
	\vect{\psi}^{(2)} & = - \frac 3 7 \nabla \Phi^{(2)} \,,                                                                              \\
	\vect{\psi}^{(3)} & =  \frac 1 3 \nabla \Phi^{(3a)}  -\frac{10}{21} \nabla \Phi^{(3b)} + \frac 1 7 \nabla \times \vect{A}^{(3c)} \,,
\end{align}
which are expressed in terms of the purely spatial functions
\begin{align}\label{eq:lpt.potentials}
	\Phi^{\,(1)} \,   & =  \varphi_{\rm ini} \,,                                                                                                   \\
	\Phi^{\,(2)} \,   & =  \frac{1}{2} \nabla^{-2}  \left[ {\Phi}^{(1)}_{,ii} {\Phi}^{(1)}_{,jj} - {\Phi}^{(1)}_{,ij}{\Phi}^{(1)}_{,ij}\right] \,, \\
	\Phi^{(3a)}\!     & =  \nabla^{-2} \left[ \det \Phi^{(1)}_{,ij}\right]\,,                                                                      \\
	\Phi^{(3b)}\!     & = \frac{1}{2}\nabla^{-2} \left[\Phi^{(2)}_{,ii}\Phi^{(1)}_{,jj}-\Phi^{(2)}_{,ij} \Phi^{(1)}_{,ij}\right] \,,               \\
	\vect{A}^{(3c)}\! & = \nabla^{-2} \left[ \nabla \Phi^{(2)}_{,i}\times \nabla \Phi^{(1)}_{,i}\right] \,.
\end{align}
where $\varphi_{\rm ini}$ is the gravitational potential $\varphi$ at $a \to 0$; explicit instructions how $\varphi_{\rm ini}$ can be obtained are given in~\autoref{sec:backscaling}. For convenience, in \aref{app:LPT}, we express these spatial functions suitably for numerical applications.

There are three simplifications in these LPT solutions worth highlighting.
First, the LPT solutions have only one degree of freedom (provided by $\varphi_{\rm ini}$), which might be surprising considering that the underlying equations are of second-order in time. Second, we ignore decaying-mode solutions, which is an additional independent assumption. Third, these LPT solutions assume an irrotational fluid motion in Eulerian coordinates; indeed only at the third order the displacement field loses its potential character, exemplified through the appearance of the vector $\vect{A}^{(3c)}$, which is actually required to maintain the zero-vorticity condition in Eulerian space.

Mathematically, these three simplifications arise from the use of the so-called slaved boundary conditions at $a=0$ on the LPT solutions~\citep[cf.][]{Brenier:2003xs,Rampf:2015}. These conditions impose initial homogeneity $\delta_{\rm m} \to 0$ where $\delta_{\rm m}$ is the density contrast,
furthermore guarantee that only one initial function needs to be provided, as well as select the purely growing-mode solutions with zero vorticity.
In the next subsection we will argue that these simplifications are seemingly intertwined with the standard procedure of generating ICs for simulations.

\subsection{Standard initial conditions for \texorpdfstring{\(N\)}{N}-body simulations}\label{sec:backscaling}

Relativistic linear Boltzmann solvers such as {\sc Camb} \citep{Lewis2000} or {\sc Class} \citep{Blas:2011} are used to evolve the coupled system of relativistic and non-relativistic species, including CDM, down to the present epoch.  Cosmological simulations are commonly performed within the Newtonian approximation and only solve for the matter species; therefore, their ICs must take into account this change in the underlying physical problem. Usually, this is handled by taking the present-day linear matter density from Boltzmann solvers and applying a suitable rescaling procedure (see below). This rescaling procedure provides a fictitious universe at initialization time with today's radiation content, is however in full agreement with relativistic perturbation theory~\citep[cf.][]{Chisari:2011,Hahn:2016roq,Fidler:2017pnb}.

Upon performing numerical simulations, \(N\)-body particles are sampled on an unperturbed lattice representing the initial homogeneity (in accordance with the slaving argument from the previous section). This initial placement of particles can be thought of happening at time zero, i.e., at $a=0$. Growing-mode initial conditions for simulations are then established by displacing the particles from $a=0$ until the time $a_{\rm start}$ when the simulation is initialized, provided we have access to the initial gravitational potential $\varphi_{\rm ini}$ at time $a=0$.

To get that initial gravitational potential, first observe that the gravitational potential is related to the matter density contrast through the Poisson equation $\nabla^2 \varphi =  \delta_{\rm m}/a$ which can be written as
\begin{equation}\label{poisson}
	\varphi =\frac{ \nabla^{-2} \delta_{\rm m}}{a} \,.
\end{equation}
From a linear Boltzmann code such as {\sc Class} or {\sc Camb}, we can obtain the linear matter density $\delta_{\rm m}^{\rm code}$, which for the rescaling procedure is assumed to be in the growing mode, i.e.,
\begin{equation}\label{deltaboltz}
	\delta_{\rm m}^{\rm code}(\vect{x},a) = C_+(\vect{x})\, D_+(a) \,.
\end{equation}
Evaluating \eqref{poisson} in the limit $a \to 0$ and expressing the matter density through~\eqref{deltaboltz}, we find
\begin{equation}\label{limit}
	\lim_{a \to 0}  \varphi =  \nabla^{-2}  C_+ \lim_{a\to0} \frac{D_+(a)}{a} \equiv \varphi_{\rm ini} \,,
\end{equation}
where we note that $D_+(a)$ is analytic around $a=0$ and can be represented as $D_+ \propto a + {\cal O}(a^4)$ for a \(\Lambda\)CDM universe.
In a final step, we express the spatial constant $C_+$ in terms of the matter density at the present \mbox{time $a_0$}, and thus obtain
\begin{equation}\label{varphiini}
	\varphi_{\rm ini} = \frac{\nabla^{-2} \delta_{\rm m}^{\rm code}(\vect{x},a_0)}{D_+(a_0)} \lim_{a\to0} \frac{D_+(a)}{a}\,.
\end{equation}
Operationally, to obtain a realization of $\varphi_{\rm ini}$, we begin with a white noise field $W(\vect{x})$ which we multiply with the total matter density transfer function $T[\delta_{\rm m}]$, an amplitude $A$ which gives the correct normalization in terms of $\sigma_8$, as well as the primordial fluctuations produced during inflation with amplitude $\sim k^{n_s/2}$. Altogether, in Fourier space for all $\|\vect{k}\|\neq0$, one thus has
\begin{equation}\label{eq:phi.ini.realization}
	\widetilde{\varphi}_{\rm ini}(\vect{k}) = A\,\widetilde{W}(\vect{k})\,k^{(n_s-4)/2} \frac{T[\delta_{\rm m}](k,a_0)}{D_+(a_0)} \lim_{a\to0} \frac{D_+(a)}{a},
\end{equation}

\noindent and  set $\widetilde{\varphi}_{\rm ini}(\text{\scriptsize $\vect{k}=\vect{0}$})=0$. Note that this is a random field, determined by $W(x)$ which creates the so-called ``cosmic variance'' whose properties are not identical to those of the ensemble average. Note that these deviations could be suppressed by the method proposed by \cite{Angulo:2016}.
However, since we will compare simulations using the same noise field, our results are already largely insensitive to cosmic variance.

\subsection{Numerical implementation of LPT}\label{sec:numerical_implementation_lpt}

The LPT potentials~\eqref{eq:lpt.potentials} needed for the 3LPT displacement can be conveniently computed in Fourier space from $\widetilde{\varphi}_{\rm ini}$ given in~Eq.\,\eqref{eq:phi.ini.realization} by using the Fast Fourier Transform (FFT).
However, some extra care should be taken for 2LPT and higher-order LPT terms:
those terms contain quadratic and higher-order non-linearities, and thus will suffer from aliasing if multiplied numerically in real space.

Aliasing leads to non-linear modes appearing at the wrong wave numbers.
For quadratic non-linearities, aliasing can be avoided by respecting Orszag's 3/2 rule~\citep{Orszag:1971}: By temporarily enlarging the computational domain in Fourier space by a factor of 3/2 per dimension while carrying out the product, the aliased modes will be all located in the padding region, which can then simply be discarded. This of course increases the memory footprint by requiring two fields of $(3/2)^3=3.375$ times the size of the original fields in 3D.

Note that the $\Phi^{(3a)}$ potential contains a cubic non-linearity, which we treat by two quadratic convolutions.
This is strictly speaking only an approximation, since for data with finite lengths, the split of a cubic convolution into quadratic convolutions is not associative, implying that some terms in the cubic convolution will be lost~\citep[e.g.][]{Roberts11}. The exact way to deal with the problem would be to evaluate cubic convolutions without such reductions, which however would require a zero padding twice as expensive per dimension compared to the 3/2 rule of the quadratic case, thereby significantly increasing the memory footprint. We find that the split into quadratic convolutions causes only a $\sim 5$\,\% error in $\Phi^{(3a)}$, which is already sub-dominant compared to $\Phi^{(3b)}$. Of course, there are more memory efficient and potentially faster implementations possible based on the in-place de-aliased convolution technique of \cite{Bowman:2011}, than what we consider here.

\begin{figure}
	\centering
	\includegraphics[type=pdf, ext=.pdf, read=.pdf, width=\columnwidth]{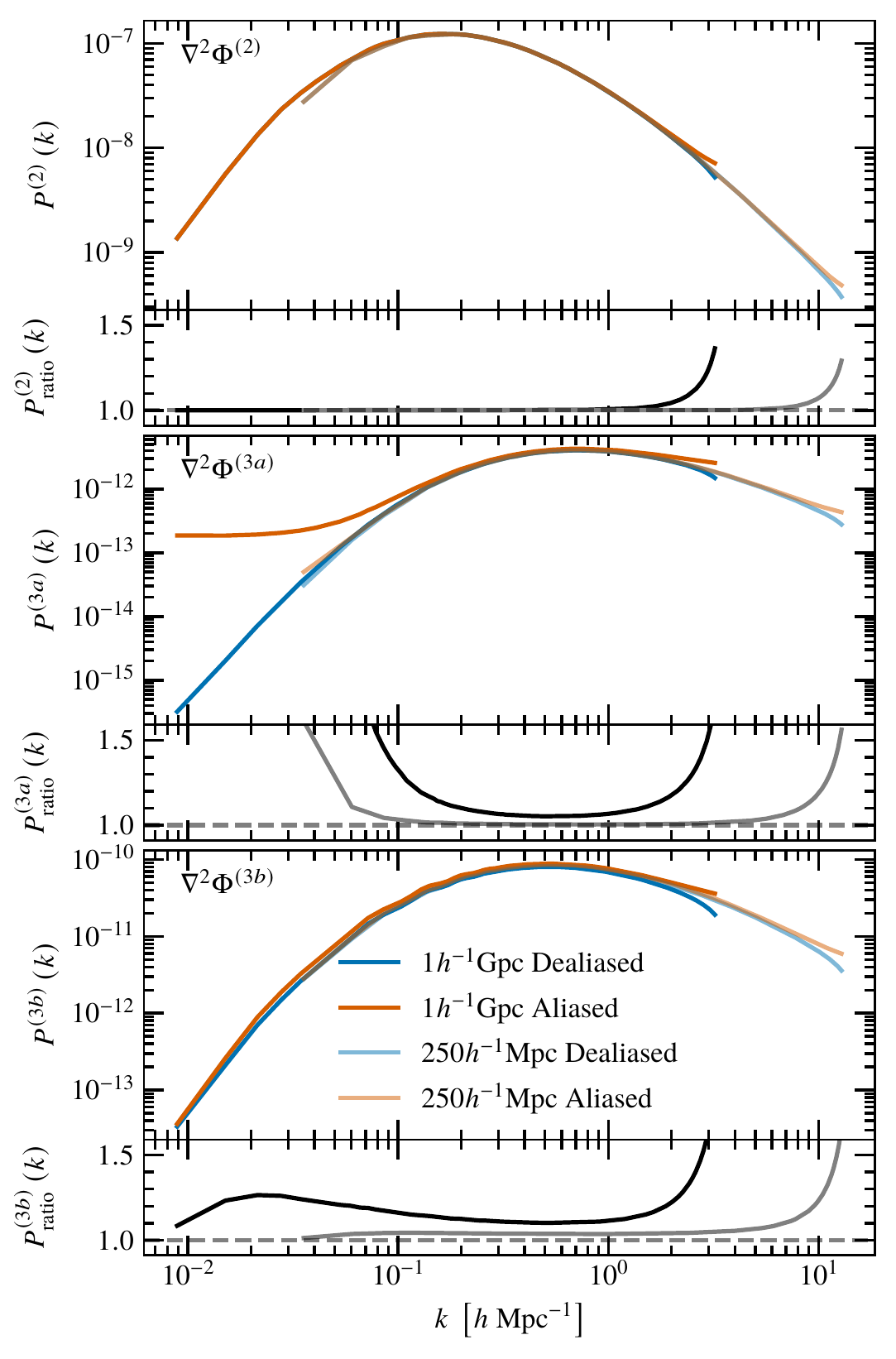}
	\caption{Power spectra of $\nabla^2\Phi^{(2)}$ (top panel), $\nabla^2 \Phi^{(3a)}$ (middle panel) and $\nabla^2\Phi^{(3b)}$ (bottom panel) for the $1\,h^{-1}{\rm Gpc}$ (darker colour) and the $250\,h^{-1}{\rm Mpc}$ box (lighter colour) with \(1024^3\) modes/particles, averaged over 50 different random realizations. We show results obtained with 3/2 padding for the de-aliasing of quadratic and cubic non-linearities (blue lines) and without de-aliasing (orange lines). Below each panel we show the ratio of the aliased to the de-aliased spectrum revealing substantial biases with significant box size/resolution dependence in the 3LPT potentials.}
	\label{fig:powerspec.orszag.rule.delta.2.3}
\end{figure}

In \autoref{fig:powerspec.orszag.rule.delta.2.3}, we show the effect of aliasing on the
power spectra of the 2LPT and 3LPT source terms $\nabla^2\Phi^{(2)}$, $\nabla^2\Phi^{(3a)}$ and $\nabla^2\Phi^{(3b)}$. We show results for two different box sizes, using \(1024^3\) modes in all cases, applying de-aliasing (blue lines) and ignoring it (red lines) for the power spectra of the ``densities'' ($\nabla^2\Phi$) of various terms in the LPT expansion.  We find that while the quadratic term (2LPT) has aliasing only at the highest wave numbers, with a relative difference of up to $30\sim 35\%$ near the particle Nyquist wave number $k_{\rm Ny} \equiv \uppi N_p^{1/3}/L_{\rm box}$,
the cubic terms (3LPT) are affected at all scales. We also find a very strong aliasing effect on the largest scales for the \((3a)\) term with a difference of nearly 3 orders of magnitude. It thus appears critical to perform de-aliasing in order to achieve a correct implementation of 3LPT. We further corroborate this aspect in \aref{sec:invariance_test}, where we demonstrate that only with de-aliasing, 3LPT shows a formal third-order convergence $\mathcal{O}(D_+^3)$, which is the expected behaviour from theory grounds.

However, even after de-aliasing, we find a weak dependence of the 3LPT terms on the chosen box size, indicated by a drop of the power close to the particle Nyquist wave number $k_{\rm Ny}$. This clearly indicates that at 3LPT we see a dependence on the UV truncation of the perturbation spectrum due to the finite resolution employed in computing the terms -- computing by de-aliased discrete Fourier transform truncates at $k_{\rm Ny}$. The coupling of modes with $k>k_{\rm Ny}$ to the modes that can be numerically represented, i.e.\ $k\le k_{\rm Ny}$, is thus missing. This effect is also present at 2LPT, but is significantly smaller. Such UV sensitivities are well-known in perturbation theory, especially within the context of loop integrations~\citep{Bernardeau:2002}, and have been discussed also in the context of the {\sc GridSPT} method of \cite{Taruya:2018}.
They are not easy to circumvent in a numerical setting since it is always expensive to increase $k_{\rm Ny}$.

We found however that, at least up to 3LPT, errors due to aliasing (and through UV sensitivities) are well below one per cent in the evolved simulations at low redshift. For more details, see the results in Appendix~\ref{appendix:aliasing}. It is thus likely unproblematic to avoid de-aliasing up to 3LPT and thereby to save memory and computing time if one is only interested in the low-redshift results of a non-linear simulation. In this paper, we carry out all simulations with fully de-aliased 2LPT and 3LPT terms. A more detailed investigation of the impact of aliasing at higher resolution, and whether including the contribution of modes $k>k_{\rm Ny}$ to the 3LPT terms could improve convergence of simulations with different mass resolution are interesting questions for future investigations.

Numerical implementations of 2LPT commonly used  in the community include the {\sc 2lptIC} software package\footnote{available from \url{https://cosmo.nyu.edu/roman/2LPT/}} introduced in~\cite{Crocce:2006}, which is based on single resolution FFTs. For multi-resolution zoom simulations, we are aware of the implementation by~\cite{Jenkins:2010}, which uses a Tree-PM approach to evaluate the 2LPT Poisson equation at higher resolution in the zoom region, and the implementation by \cite{Hahn:2011} in {\sc Music}\footnote{available from \url{https://bitbucket.org/ohahn/music}}, which uses a combined algebraic multigrid and FFT approach. Version~4 of the {\sc Pinocchio} code \citep{Munari:2017} implements the longitudinal part of 3LPT to determine the large-scale clustering of haloes in the rapid mock catalogue {\sc Pinocchio} scheme \citep{Monaco:2002}. To our knowledge, there exists no publicly available implementation of 3LPT \(N\)-body initial conditions that includes transversal modes and performs a correct de-aliasing of higher-order terms, and which could therefore allow an accurate assessment of truncation errors and transients.

We make our implementation publicly available as the {\sc Music2-MonofonIC} software package\footnote{available from \url{https://bitbucket.org/ohahn/monofonic}}  that comprises a distributed memory parallelized (MPI+threads) implementation of all algorithms discussed here. It is, however, currently restricted to single resolution (mono-grid) simulations. {\sc Music2} (Hahn et~al.~2020, in prep.) will be the next update to the {\sc Music} software package of~\cite{Hahn:2011}, which will at a later stage also support zoom simulations.

%%%%%%%%%%%%%%%
\subsection{Initial particle placements and particle linear theory}\label{sec:lattice_PLT}

As mentioned earlier, the creation of ICs require as input a homogeneous and isotropic particle distribution. In practice, this state is commonly realized by a simple cubic (SC) lattice, where particles are arranged on a regular grid. Note that other alternatives for  homogeneous non-regular particle distributions have been proposed \citep[e.g.][for `glass', `quaquaversal', and `CCVT' particle distributions, respectively]{White:1996,Hansen:2007,Liao:2018}.

The case of an SC lattice is particularly convenient since one can simply have one particle per Fourier mode so that the SC lattice coincides with the FFT mesh used to compute the LPT terms. An important drawback is that Fourier modes on such lattices do not grow identically as in the continuum fluid limit owing to self-interactions of the discrete lattice, as has been pointed out in a series of papers \citep{Joyce:2005,Marcos:2006cn,Joyce:2007}.

These discreteness effects are present in all particle distributions (but harder to quantify in non-regular lattices). This is a consequence of the gravitational softening length being smaller than the mean inter particle separation \cite[see also][where the effect is much more dramatic in the case of multiple particle species]{Angulo:2013}. The effect is, of course, strongest at early times when the lattice is still close to regular and physical density fluctuations are still small. \cite{Joyce:2005} and \cite{Marcos:2008} have shown how the deviation from the fluid case can be calculated for early times in ``particle linear theory'' (PLT). Recently, this work has been extended by \cite{Garrison:2016} who use the earlier solutions to compensate the initial particle displacements and velocities by the expected discreteness of the lattice at linear order and for early times.

To quantify the impact of particle discreteness, we have implemented an {\it optional} PLT correction in our initial conditions. Note that our version is similar, but deviates in some respects from that of \cite{Garrison:2016} (e.g. avoiding their artificial boosting of the counter-PLT modes). In \aref{sec:plt}, we provide details on our implementation. Nonetheless, we remark that the PLT correction is a decaying mode and thus vanishes over the course of the simulation. In principle, this can be avoided by artificially boosting PLT corrections, as proposed by \citealt{Garrison:2016}, or by repeatedly applying it over time -- but note that PLT is only valid at linear order and currently there is no higher-order theory available, which limits its applicability at late times.

Our preferred choice for most parts of this paper -- instead of PLT -- is to use a face-centered-cubic (FCC) lattice constructed by shifting four SC lattices,\footnote{We have also considered body-centered-cubic lattices (half the particle load compared to FCC), leading to results between SC and FCC lattices.}
	and imposing the displacement and velocity using a corresponding shift of the respective fields using a Fourier shift with the FFT.  The FCC lattice is more isotropic than the SC lattice, with four times as many particles as our standard simulations. In fact, \cite{Marcos:2008} showed that Bravais lattices other than SC exhibit weaker deviations from the fluid limit at a fixed particle number, due to higher symmetries \citep[cf. also][]{Stuecker:2020}.

\subsection{Validation}\label{sec:validation}

\begin{figure}
	\centering
	\includegraphics[type=pdf, ext=.pdf, read=.pdf, width=0.98\columnwidth]{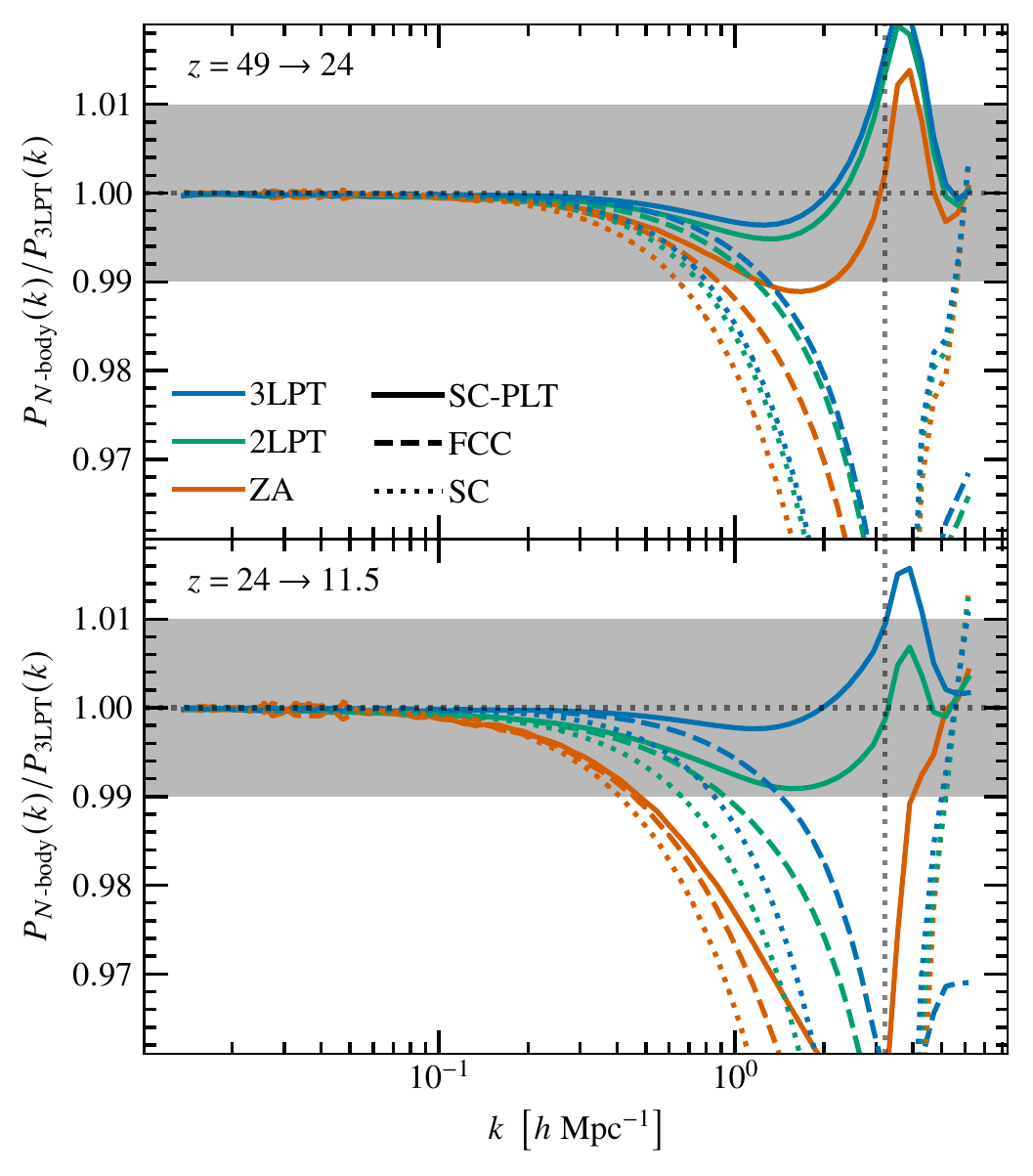}
	\caption{Top (bottom) panel: ratio between power spectra from \(N\)-body simulations and from 3LPT at \(z = 24\) (\(11.5\)). The \(N\)-body simulations were evolved from \(\zstart = 49\) (\(24\)) down to \(z = 24\) (\(11.5\)) using initial conditions based on ZA, 2 and 3LPT (blue, green, and orange, respectively) and initialized from different perturbed lattices. Solid lines show the results of the SC lattice initial conditions including corrections from particle linear theory (PLT; see~\autoref{sec:lattice_PLT}, to correct for particle discreteness), and the dotted lines without. Dashed lines show the result of the particle oversampling using an FCC lattice without PLT correction. The vertical dotted line indicates the particle Nyquist wave number of the initial conditions. One percent agreement is represented as a shaded area and a perfect agreement as a dotted horizontal line. With PLT the simulations agree at per cent level with 3LPT up to the particle Nyquist wave number when 2LPT or 3LPT is used.}
	\label{fig:powerspec.plt.sc.convergence}
\end{figure}

In this subsection we present a validation of our numerical and analytic tools by comparing perturbative and numerical solutions, emphasizing the role of particle discreteness.

In \autoref{fig:powerspec.plt.sc.convergence}, we show the ratios of various numerical predictions for the non-linear power spectrum to that in LPT. Specifically, we employ an \(N\)-body simulation with $N_p = 1024^3$ particles in a $L_{\rm box}= 1 h^{-1}$Gpc box, and compute the LPT fields on a grid with $1024^3$ points.
Coloured lines display the case where the initial conditions were computed using 3rd, 2nd or 1st-order LPT. Solid and dotted lines indicate cases where PLT corrections have been included or not, while dashed lines refer to using an FCC lattice instead of PLT. In the top panel we start our simulations at $z=49$ and evolve them until $z=24$, whereas in the bottom panel we start at $z=24$ and evolve until $z=11.5$.

As is evident from the figure, without PLT corrections, there is a clear discrepancy between perturbative and numerical solutions, with the latter showing a significant loss of power at the particle Nyquist wave number of the particle lattice, $k_{\rm Ny} \equiv \uppi N_p^{1/3}/L_{\rm box}$. The loss of power is less prominent when using an FCC lattice. Note that this power suppression is almost insensitive to the order of LPT used to generate the ICs, which could create the illusion of proper convergence in simulation results. This is, however, convergence to the discrete solution and not to the fluid solution.

In contrast, there is a remarkable agreement between the \(N\)-body and perturbative solutions when PLT is enabled. Specifically, there is a 1\% agreement between 2LPT, 3LPT, and the numerical simulation up to $k_{\rm Ny}$ (indicated by vertical dotted line). Note that only with PLT corrections, one is able to see the improvement brought by higher order ICs, but even ZA performs well  in this test if initialized early enough ($\zstart=49$).

\begin{figure}
	\centering
	\includegraphics[type=pdf, ext=.pdf, read=.pdf, width=0.98\columnwidth]{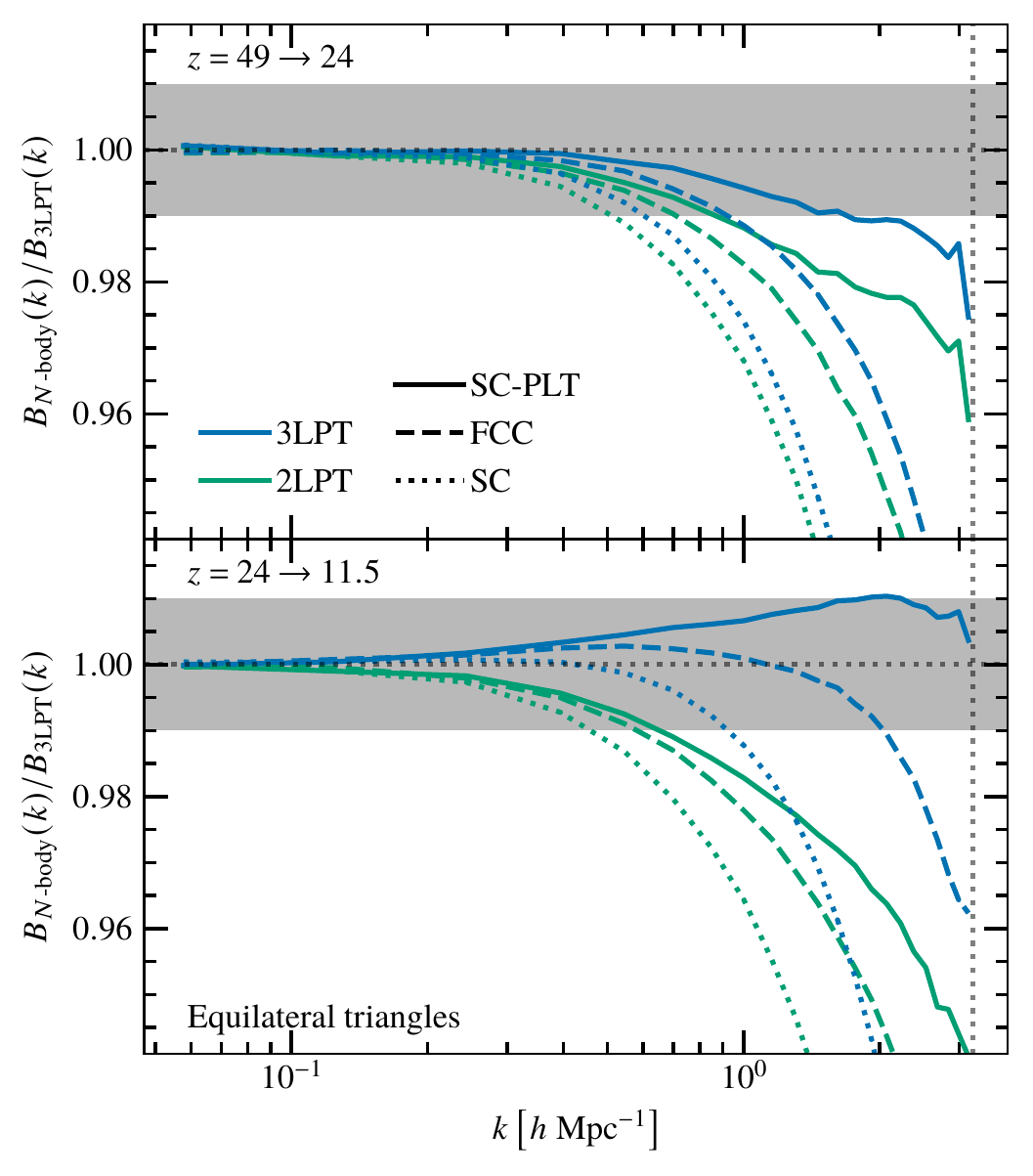}
	\caption{Same as \autoref{fig:powerspec.plt.sc.convergence}, but now showing the ratio between the respective equilateral bispectra at $z=24$ ($11.5$) in the top (bottom) panel. Only the simulations generated with 3LPT and with PLT corrections lead to $\sim1$~per cent agreement between simulations and 3LPT at all $k$ below the particle Nyquist wave number, while 2LPT undershoots already at much smaller $k$. ZA is not even present on this figure because it is off at all scales (ratio between 0.8 and 0.5).}
	\label{fig:bispec.equilateral.plt}
\end{figure}

In \autoref{fig:bispec.equilateral.plt} we show the same figure as before (\autoref{fig:powerspec.plt.sc.convergence}) but now for the equilateral bispectrum (i.e., with $k=k_1=k_2=k_3$; see Eq.\,\eqref{eq:bispectrum} for the used bispectrum definition). As for the power spectrum, we find a systematic power deficit in the numerical solutions when PLT is not considered, and an apparent convergence to the wrong solution. Note that also here the loss of power is less pronounced when using an FCC lattice.

With PLT corrections, a good agreement between numerical and perturbative bispectra is achieved, but only for the higher-order version of the ICs. Unlike for the power spectrum, ZA ICs perform very poorly in this test. On large scales, they underestimate the bispectrum at about \(20\%\), and up to \(50\%\) on small scales, regardless of PLT corrections. Thus, it is not shown in the figure. This suggests that ZA ICs have a leading-order transient in the bispectrum and should {\it not} be used to set up simulations if high precision is required.

Note that for 3LPT with PLT correction, the error between our two redshift ranges changes sign, overshooting in the redshift range $z=24 \to 11.1$ (lower panel in~\autoref{fig:powerspec.plt.sc.convergence}) while undershooting for $z=49 \to 24$ (upper panel in~\autoref{fig:powerspec.plt.sc.convergence}).
From a theoretical point of view, the next-to-leading (one-loop) correction to the bispectrum involves density correlators up to fourth order (the ``411'' contribution) which should be of the same magnitude as those that we implicitly determine on the grid points using 3LPT ICs. Thus, 3LPT might not be sufficiently converged for predicting the nonlinear bispectrum at the few percent level, and it is likely that 4LPT might be required for an accurate comparison with simulations.\footnote{This should be contrasted to the convergence studies related to the power spectrum (\autoref{fig:powerspec.plt.sc.convergence}) where 4LPT becomes relevant only at 2-loop and beyond.} We can then speculate that this overshooting might imply that a starting redshift of 11.5 is already too late for 3LPT. We remark that 2LPT at lower redshift becomes worse than 3LPT without PLT correction or with FCC oversampling, even if the sign does not change, thus suggesting that 2LPT suffers more from the lack of some higher-order correction for the bispectrum. The impact of 4LPT ICs on the bispectrum will be investigated in future works.

From the previous plots, the impact of particle discreteness is obvious. Unfortunately, PLT corrections work for a relatively short period if not artificially boosted (as \citealt{Garrison:2016} propose)
and/or multiply applied at various stages.
Furthermore, the perturbative estimates to the discrete lattice effects have only been computed to first order.
For this reason our reference simulation will not be with the PLT corrections switched on. Instead, we use a FCC lattice as described at the end of \autoref{sec:lattice_PLT}.

Since the Fourier perturbation modes are initially specified on the SC lattice, we oversample the Fourier modes and phase-shift (in Fourier space) the $n$LPT displacements to obtain them in the FCC case. This enables us to compare how the exact same mode spectrum converges on various statistics as we change the particle number and degree of isotropy of the underlying particle lattice onto which these modes are imposed. In particular, the SC lattice samples 1:1 the Fourier modes, meaning that we have one particle per mode in an SC simulation, and perturbations at the particle Nyquist wave number are basically represented by two particles. By oversampling, no new perturbation modes are added, which is in contrast to usual convergence studies carried out for cosmological simulations, where with an increase in the particle number also new perturbations are added. When we compare an SC against an FCC simulation, we thus test convergence at fixed initial modes, investigating only the impact of particle sampling. Keeping the modes fixed is especially important since, as we will show in~\autoref{sec:convergence_radius}, increasing the particle Nyquist mode of sampled fluctuations increases the variance of the fluctuations realized in the box, and in turn influences the degree of non-linearity at the starting time.
In contrast, allowing the ICs to change would render an accurate comparison of starting redshifts impossible. A somewhat orthogonal test of the impact of particle discreteness comparing the different lattices, as well as also `glass' \citep{White:1996} and other \citep[e.g.][]{Hansen:2007,Liao:2018} pre-initial conditions, at fixed particle number would be a very interesting project for a future study. We found however that glass pre-initial conditions (results not shown in this paper) did not improve the convergence of power and bispectrum over those presented for the perturbed SC lattice below in \autoref{sec:results}. This shows that indeed the small-scale interactions between \(N\)-body particles are responsible for the discreteness errors, rather than the particular anisotropies of a given  pre-initial conditions particle distribution.

In summary, the results presented above clearly indicate that, without discreteness corrections, \(N\)-body simulations deviate strongly from the fluid limit during the perturbative phase. This effect is larger, the earlier the simulation is initialized. Therefore, to minimize these errors, one should delay the initialization time to the latest possible moment. In the following section, we will formally investigate what this ``latest possible moment'' is in the context of~LPT.

%%%%%%%%%%%%%%%%%%%%%%%%%%%%%%%%%%%%%%%%%%%%%%%%%%
\section{Convergence radius of LPT and starting time for simulations}\label{sec:convergence_radius}

In the previous section we argued that early starting redshifts for simulations are accompanied by significant discreteness effects. These effects decrease in magnitude at lower redshifts, thus the latest possible starting redshift is desirable. On the other hand, LPT ICs are based on a single-stream fluid description implying that once particle trajectories cross for the first time (``shell-crossing''), the single-stream fluid equations and thus LPT become invalid.

In this section we will discuss several aspects regarding determining the point at which LPT breaks down, which assist clarifying the appropriate redshift-window for generating initial conditions. Necessary definitions related to the LPT series are provided in~\autoref{sec:LPTradius}, while we outline numerical tests for estimating the convergence radius in~\autoref{sec:ratiotest}. Fairly complementary to those numerical tests, there are also ways to estimate the convergence radius directly from theory; the main ideas are sketched in~\autoref{sec:theorybound} while further technical details are provided in Appendix~\ref{app:theoryconvergence}.

\subsection{LPT series and its convergence radius}\label{sec:LPTradius}
The breakdown of LPT is intimately linked to finding the radius of convergence of the Taylor series of the displacement
\begin{equation}
	\vect{\psi}  = \sum_{n=1}^\infty  \vect{\psi}^{\left(n\right)} \left(\vect{q}\right)  D_{+}^{n} \,.
	\label{eq:displacement_taylor_series-rep}
\end{equation}
Indeed, as it is known from complex analysis, the radius of convergence, $R_{\rm conv}$, of any Taylor series is limited by the nearest singularity in the complex domain of its argument. For the LPT series, which is a time-Taylor series with time variable $D_+$, the nearest singularity could be in the real domain of $D_+$ but could  also take complex values.\footnote{To illustrate the argument of singularities, let us consider a toy example. Let us assume that the exact, non-perturbative displacement is $\psi(D_+) = 1/(1+D_+^2)$, which has complex singularities at $D_+= \pm {\rm i}$. In LPT, we represent this displacement by a Taylor series, i.e., $\psi= \sum_n c_n D_+^n$, but because of the appearance of these two complex singularities, we can evolve particles only for $0 < |D_+| < 1$, i.e., within the disc of convergence. If shell-crossing has not occurred until $|D_+|=1$, then we should seek for an analytic continuation technique, like the one of Weierstra\ss, that in our case amounts to evolve particles until $|D_+| = D_< <1$, then re-expand and determine the new Taylor coefficients around $D_<$. The series around $D_<$ will generally have a new radius of convergence that allow us to continue and follow the particles for $|D_+|> D_<$, possibly involving repetitive re-expansions, until a real singularity in time and/or shell-crossing, occurs.}\@
For example, a real singularity could appear at shell-crossing -- where the density becomes formally infinite -- which in any case marks the break down of LPT.

Regardless of the precise nature of the singularities, a single-time push-forward displacement of particles (as done for the IC creation) is only meaningful mathematically as long as the chosen time step is within the disc of convergence  spanned by $R_{\rm conv}$ \citep[cf.][]{Rampf:2015}:
\begin{equation}\label{eq:maxtimestep}
	\left| D_+ (a_{\rm max}) \right| < R_{\rm conv} \,,
\end{equation}
Hence $a_{\rm max}$ sets formally the maximal scale-factor for using LPT initial conditions for a simulation. Note that typically, \(N\)-body simulations adopt an heuristic criterion requiring the average amplitude of fluctuations $\sigma$ at the resolution scale to be small, i.e., \(\sigma\ll 1\).
While such empirical criteria are certainly useful, we suggest here to take also the breakdown of LPT into account.

In the following subsection we provide two complementary methods to estimate $R_{\rm conv}$: the numerical ratio test, see \autoref{sec:ratiotest}, and a fully analytical method that exploits a theoretical lower bound on $R_{\rm conv}$, as outlined in \autoref{sec:theorybound}.
These estimates translate into threshold for the latest possible initialization time of the simulation (based on theory grounds), which is discussed in~\autoref{sec:conv_numerical}.

%%%%%%%%%%%%%%%%%%%%%%%%%%%%%%%%%%%%%%%%%%%%%%%%%%

\begin{figure}
	\centering
	\includegraphics[type=pdf, ext=.pdf, read=.pdf, width=0.98\columnwidth]{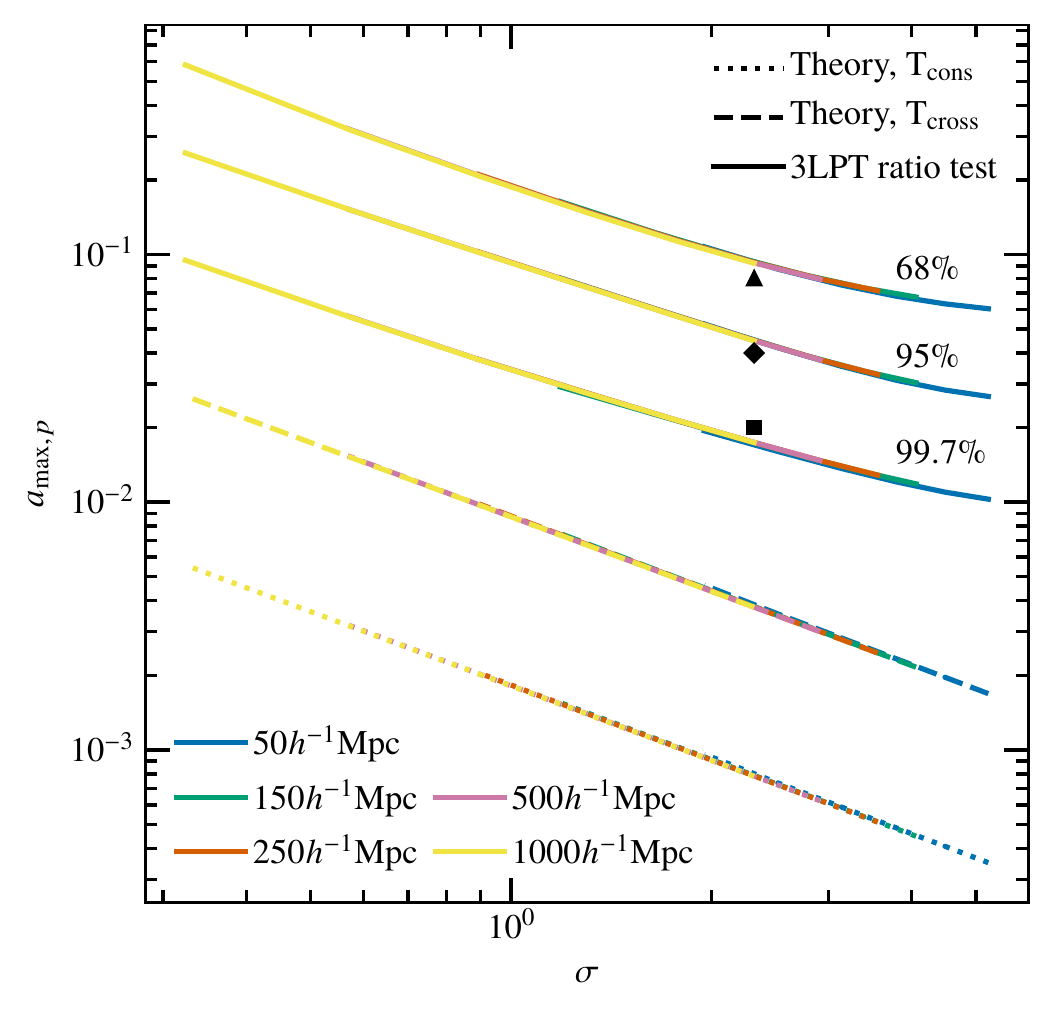}
	\caption{Maximum scale-factor \(a_{\max,p}\) for which various methods predict convergence of LPT, thereby pinning down the latest possible time when simulations should be  initialized. Specifically, we show \(a_{\max,p}\) as a function of the fluctuation scale \(\sigma\) for different box sizes \(L_{\rm box}\) (colors), averaged over 10 random realizations.
	The solid lines are the result of the numerical ratio test using our implementation of 3LPT, the dashed and dotted lines are analytical bounds on the radius of convergence computed using~\eqref{eq:dplus.theory} for \({\rm T}= \mathrm{T}_{\mathrm{cons}} = 0.022\) (dotted) and \({\rm T} =\mathrm{T}_{\mathrm{cross}} = 0.107\) (dashed). The values of $p$ denote the percentage of particle displacements in the simulation that satisfy the LPT convergence condition~\eqref{eq:maxtimestep}. We chose here the percentiles corresponding to the 1, 2 and \(3\sigma\) (light to bright) standard deviation of the normal distribution. For the theoretical bounds, we use \(p=99.9\%\).
	Our simulations (box length \(1h^{-1}\)Gpc) are indicated in the figure with a square symbol for \(z_{\mathrm{start}} = 49\), a diamond for \(z_{\mathrm{start}} = 24\) and a triangle for \(z_{\mathrm{start}} = 11.5\).}
	\label{fig:convergence_sigma_amax}
\end{figure}

\subsection{Numerical estimation of the radius of convergence}\label{sec:ratiotest}

A particularly simple method for estimating $R_{\rm conv}$ is to consider the~$L^2$ norm of the displacement~\eqref{eq:displacement_taylor_series-rep}, i.e.,
\begin{equation}
	\Psi \equiv \sum_{n=1}^\infty  \left\| \vect{\psi}^{\left(n\right)} \left(\vect{q}\right)  \right\| D_{+}^{n} \,,\label{normedLPTseries}
\end{equation}
and perform the convergence tests for that series \citep[cf.][]{PODVIGINA2016320}. For this, we use the ratio test which states that the radius of convergence $R_{\rm conv}$ of the series is
\begin{equation}
	\frac{1}{R_{\rm conv}} = \lim_{n \to \infty} \text{\LARGE $\sfrac{\text{\normalsize $\big\| \vect{\psi}^{(n)}  \big\|$}\,}{\,\text{\normalsize $\big\| \vect{\psi}^{(n-1)}\big\|$}}$}\label{eq:ratiotest}
\end{equation}
(if that limit exists). In all generality, it is thus the large-$n$ limit of Taylor coefficients that decides questions about convergence.
Of course, in numerical implementations of the ratio test, the actual limit $n \to \infty$ can only be reached by employing numerical extrapolation methods (see next paragraph), which in principle can be performed to very high accuracy.
Nonetheless, we remark that in the present case where we have numerically implemented perturbative solutions only up to third order, the resulting numerical extrapolation is fairly crude (see the following section for a more rigorous yet more restrictive method).

\cite{DombSykes:1957} introduced a numerical extrapolation method in a non-cosmological context. There, one draws the  plot $\| \vect{\psi}^{(n)} \| / \| \vect{\psi}^{(n-1)} \|$ versus $1/n$, and takes the $y$-intercept (``$n \to \infty$'') as the estimate for $1/R_{\rm conv}$. For solutions until third order, we have two such ratios, leading to the two tuples
\begin{equation}
	\left(\,\text{\LARGE $\sfrac{\text{\normalsize $1$\!}}{\text{\normalsize $2$}}$}\,, \,\text{\LARGE $\sfrac{\| \vect{\psi}^{(2)} \|}{ \| \vect{\psi}^{(1)} \|}$}\, \right)
	\quad\textrm{and}\quad
	\left(\,\text{\LARGE $\sfrac{\text{\normalsize $1$}\!}{\text{\normalsize $3$}}$}\,, \, \text{\LARGE   $\sfrac{\|\vect{\psi}^{(3)} \|}{\| \vect{\psi}^{(2)} \|}$} \,\right),
\end{equation}
from which we can perform a simple linear extrapolation to the~$y$-intercept. This argument leads to the following estimate for the radius of convergence of the LPT series,
\begin{equation}
	\frac{1}{R_{\rm conv}\left(\vect{q}\right)} \approx  3\text{\LARGE $\sfrac{\text{\normalsize $\big\| \vect{\psi}^{(3)}  \big\|$}\,}{\,\text{\normalsize $\big\| \vect{\psi}^{(2)}\big\|$}}$}
	-2  \text{\LARGE $\sfrac{\text{\normalsize $\big\| \vect{\psi}^{(2)}  \big\|$}\,}{\,\text{\normalsize $\big\| \vect{\psi}^{(1)}\big\|$}}$} \,.
\end{equation}
Note the somewhat artificial dependence of $\vect{q}$ in $R_{\rm conv}$ which we have included here by hand; the actual radius of convergence of LPT is obtained by searching for the global minimum $R_{\rm actual} = \min_{\vect{q}} R_{\rm conv} (\vect{q})$, which is, so to say, the worst-case scenario over the whole spatial domain (in the present case: in the simulation box).
However, since we employ random initial conditions,
we expect some points in the realization of this probability distribution to have extreme values.
For example, if $\| \vect{\psi}^{(2)} \| \simeq 0$ locally (indicating local pancake formation) but then with $\| \vect{\psi}^{(3)} \|  = {\cal O}(1)$ (non-negligible gravitational couplings from small spatial scales), then our low-order convergence test will predict a tiny radius of convergence. We stress however, that in such collapse scenarios higher-order ratios are likely to change the outcome of the convergence test significantly. We will come back to this issue in a forthcoming paper.

Therefore, we expect to observe outlier values which should be disregarded.
For that reason we choose for the ratio test a statistical approach as outlined next. See the following section for a method that is only sensitive to the deterministic radius of convergence.

Inverting \(\left| D_{+} \left( a_{\max} \right) \right| \lessapprox R_{\rm conv}\) gives the upper limit \(a_{\max}\)  (or the lower limit \(z_{\min}\)) on the starting time of the simulation, i.e.,  \(\astart \lessapprox a_{\max}\). We can compute \(a_{\max}\left( \vect{q} \right)\) for all \(\vect{q}\) in the simulation box and get the value of \(a_{\max,p}\) at some chosen percentile \(p\). This means that by taking \(\astart = a_{\max,p}\), \(p\) percent of the points will fulfil the condition~\(\left| D_{+} \left( a_{\max} \right) \right| \lessapprox R_{\rm conv}\).
Results of the ratio test are shown in \autoref{fig:convergence_sigma_amax} and discussed in \autoref{sec:conv_numerical}.

%%%%%%%%%%%%%%%%%%%%%%%%%%%%%%%%%%%%%%%%%%%%%%%%%%
\subsection{Analytical bound on the radius of convergence}\label{sec:theorybound}

Complementary to the above numerical method,
\cite{Zheligovsky:2014} and \cite{Rampf:2015} have shown that, being equipped with explicit all-order recursion relations for LPT, it is also  possible to obtain an {\it analytical bound} $D_+^{\rm theory}$ {\it on the radius of convergence} $R_{\rm conv}$, i.e.,
\begin{equation}
	|D_+^{\rm theory}| < R_{\rm conv} \,.
\end{equation}
This bound comes actually from \mbox{``order \(\infty\)''} in LPT (see Appendix~\ref{app:theoryconvergence}), but at the same time, since only very weak assumptions on the initial conditions are imposed, leads to a bound that is (much) smaller than \(R_{\rm conv}\).
Nonetheless, this analytical bound is to some extent complementary to the low-order estimate as employed in the previous section, and therefore included in our studies.
The numerical exploitation of all this and the ratio test will be done in the following section.

In Appendix~\ref{app:theoryconvergence} we outline three ways how theoretical bounds on the radius of convergence can be estimated. For the present work we use the two methods that lead to the best bounds, namely case (a) where the employed bounds are conservative (``cons'') but already slightly improved in comparison to those of \cite{Rampf:2015}, and (b) a new -- however very optimistic -- method where we ignore complex time singularities and, possibly, can increase the bound until shell-crossing (``cross''). We find
\begin{equation}
	\label{eq:dplus.theory}
	D_+^{\rm theory} = \frac{\rm T}{\| \nabla_i \nabla_j \varphi_{\rm ini} \|} \,,
\end{equation}
where in case (a) we have \({\rm T} = {\rm T}_{\rm cons} = 0.022\) and in case (b) we have \({\rm T} = {\rm T}_{\rm cross} = 0.107\).
In~\autoref{fig:convergence_sigma_amax} we show how these theoretical bounds relate to constraints for the allowed scale-factor where simulations can be  initialized. As expected, these theoretical bounds are much weaker than the one coming from the ratio test. We note that for the theoretical bounds we have fixed  \(p=99.9\%\) in~\autoref{fig:convergence_sigma_amax} to minimize an unwanted dependence on the box size.

%%%%%%%%%%%%%%%%%%%%%%%%%%%%%%%%%%%%%%%%%%%%%%%%%%
\subsection{Upper limits on initialization time for simulations}\label{sec:conv_numerical}

The analytical bound as well as the complementary numerical ratio test translate into upper limits on the initialization time $\astart \lessapprox a_{\max}$ for the simulations. In~\autoref{fig:convergence_sigma_amax} we summarize the resulting upper limits as a function of the fluctuation scale $\sigma$. Evidently, the numerical tests provide much stronger bounds on $a_{\rm max}$ than the analytical method, however we remark again that the employed numerical tests only go to third order in PT -- which only provides two data points for the numerical extrapolation. Thus, these stronger bounds, which could change at higher orders, should still be read with caution. The analytical method, by contrast, provides bounds that are rock-solid.

For sufficiently small $\sigma$, the results from the numerical ratio test are approximately straight lines using log-log scaling. Using this as a working assumption, we can give some simple formula that could help determine quickly whether a given starting time is inside or outside the radius of convergence, and to choose the optimal one for a simulation.

For the ratio test, the maximum starting time is well approximated by:
\begin{equation}
	\label{eq:ratio.test.fit}
	a_{\max, n} \approx \frac{0.2}{n} \sigma^{-0.8},
\end{equation}
where $\sigma$ is the standard deviation, and $n$ is the number of standard deviations required to be inside the convergence radius (since the distribution is not Gaussian, we map this to the corresponding percentiles, i.e.\ $p=68$, 95, and 99.7 for $n=1$,2, and 3, respectively).
We remark that the value of \(\sigma\) can be calculated on the random realization prior to the back-scaling step and thus, the initialization time for the simulation can be chosen at the last moment from Eq.\,\eqref{eq:ratio.test.fit} by specifying the number of standard deviations \(n\).

In contrast, the theoretical bounds from the previous section are much closer to
\begin{equation}
	a_{\rm max} \approx \frac{\rm T}{12.2 \sigma} \,,
\end{equation}
where \({\rm T}\) is either \({\rm T}_{\rm cons}\) or \({\rm T}_{\rm cross}\) as defined in the previous section.
The scaling  \(\propto\sigma^{-1}\) corresponds to what one would expect from linear theory structure growth in an Einstein--de Sitter (EdS) universe, such that these bounds basically require that the fluctuations do not exceed a fixed amplitude at the starting time.

The theoretical bounds are thus much more conservative than the numerical estimates we obtained from the ratio test: the former suggest starting times \(z_{\rm start}\gtrsim 100\), while the latter suggest this could be a factor \(\gtrsim5\) lower, provided that we exclude certain outliers (due to the employed low-order numerical test and the nature of random ICs). How these numbers translate into a given precision for the summary statistics at late times is a non-trivial question due to the increased impact of discreteness errors for early starts that we have discussed before. We will attempt to disentangle the two in the next sections.

%%%%%%%%%%%%%%%%%%%%%%%%%%%%%%%%%%%%%%%%%%%%%%%%%%
\section{Non-linear Simulations and Analysis}\label{sec:numerical_results}

Here we first describe our cosmological numerical simulations (\S\ref{sec:numeral_evolution}) and then the methods we employ to quantify the properties of the respective non-linear density fields (\S\ref{sec:statistic}).

%%%%%%%%%%%%%%%%%%%%%%%%%%%%%%%%%%%%%%%%%%%%%%%%%%%%%%%%%%%%
\subsection{Numerical evolution}\label{sec:numeral_evolution}
\begin{table}
	\begin{center}
		\begin{tabular}{ll}
			\hline
			\texttt{Softening}              & 0.016276 \\
			\texttt{ErrTolForceAcc}         & 0.002    \\
			\texttt{ErrTolIntAccuracy}      & 0.025    \\
			\texttt{ErrTolTheta}            & 0.5      \\
			\texttt{MaxRMSDisplacementFac}  & 0.25     \\
			\texttt{MaxSizeTimestep}        & 0.01     \\
			\texttt{TypeOfOpeningCriterion} & 1        \\
			\hline
		\end{tabular}
	\end{center}
	\caption{Force and time integration accuracy parameters of {\sc L-Gadget3} employed for the simulations.}
	\label{tab:lg3-param}
\end{table}

For all numerical results, we use the {\sc L-Gadget3} code \citep{Angulo:2012} which is a heavily modified and optimized version of the {\sc Gadget-2}  \(N\)-body code \citep{Springel:2005}. It implements a parallel tree-PM method with periodic boundary conditions. For all simulations, we employ a $2048^3$ PM grid, with the force and time integration parameters listed in \autoref{tab:lg3-param}. The values of those parameters are expected to yield sub-percent convergence in the low-redshift power spectrum on scales $k \lesssim 2\ihMpc$ \citep[see Fig.\ 2 of][]{Angulo:2020}.

\begin{table}
	\begin{center}
		\begin{tabular}{ccccl}
			\hline
			PT   & $z_{\rm start}$ & lattice & $m_p/10^{10}h^{-1}{\rm M}_\odot$ \\
			\hline
			ZA   & 49              & SC      & 8.04                             \\
			ZA   & 99              & SC      & 8.04                             \\
			ZA   & 199             & SC      & 8.04                             \\
			\hline
			2LPT & 11.5            & SC      & 8.04                             \\
			2LPT & 24              & SC      & 8.04                             \\
			2LPT & 49              & SC      & 8.04                             \\
			2LPT & 99              & SC      & 8.04                             \\
			2LPT & 199             & SC      & 8.04                             \\
			\hline
			3LPT & 11.5            & SC      & 8.04                             \\
			3LPT & 11.5            & FCC     & 2.01                             \\
			3LPT & 24              & SC      & 8.04                             \\
			3LPT & 24              & FCC     & 2.01                             \\
			3LPT & 49              & SC      & 8.04                             \\
			3LPT & 49              & FCC     & 2.01                             \\
			\hline
		\end{tabular}
	\end{center}
	\caption{\label{tab:simulations} List of the simulations we use in this paper, differing in the order of the LPT used to initialize them as well as the starting time. All simulate the same $L_{\rm box}=1\,h^{-1}{\rm Gpc}$ box with the cosmology described in \autoref{sec:numeral_evolution}. The `SC' runs have $1024^3$ particles in an SC lattice, the `FCC' runs $4\times 1024^3$ particles in an FCC lattice. All simulations sample the same $1024^3$ modes.}
\end{table}

We use the following cosmological parameters, consistent with the {\sc Planck2018}+LSS results \citep{Planck:2020}: $\Omega_{\rm m} = 0.3111$, $\Omega_\Lambda = 0.6889$, $\Omega_{\rm b}  = 0.04897$, $h = 0.6766$, $\sigma_8 = 0.8102$ and $n_s = 0.9665$. The transfer function has been calculated using the {\sc Class} code\footnote{available from \url{http://class-code.net/}} \citep{Blas:2011} at $z=0$, and then scaled back using the linear theory growth factor $D_+(z)$ to the various starting redshifts we use (see discussion in \autoref{sec:backscaling}).

For the simulations in this study, we use a box of linear size $L_{\rm box} = 1 h^{-1}$\,Gpc, and use $N_{\rm part} = 1024^3$ whenever we initialize particles on a simple cubic (SC) lattice. This yields an \(N\)-body particle mass of  $m_{\rm part} = 8.04 \times 10^{10}\,h^{-1}\solarmass$. In some cases, we oversample the white noise fields with different lattices (see discussion in \autoref{sec:lattice_PLT}).
When an FCC lattice is used, the number of particles is $4\times1024^{3}$ and the mass $2.01 \times 10^{10}\,h^{-1}\solarmass$.
For each of these simulations, we have carried out versions employing de-aliased (c.f.\S\ref{sec:numerical_implementation_lpt}) ZA, 2LPT and 3LPT initial conditions. For testing, we have also run versions with 2LPT and 3LPT ICs without dealiasing
(see Appendix~\ref{appendix:aliasing}.) In \autoref{tab:simulations} we list all simulations we use in our paper.

%%%%%%%%%%%%%%%%%%%%%%%%%%%%%%%%%%%%%%%%%%%%%%%%%%%%%%%%%%%%
\subsection{Statistical analysis of density field data}\label{sec:statistic}
We briefly summarize below the various quantities that we will investigate in \autoref{sec:results} along with their respective definitions and the way we computed/evaluated them. Throughout, we use cloud-in-cell interpolation \citep[CIC; cf.][]{Hockney:1981} to deposit the \(N\)-body particles to a regular grid. From this density field, we compute one-, two- and three-point statistics.

\subsubsection{Density field and one-point statistics}\label{subsubsec:density.one.point.stats}

To study the distribution function of densities, we use a grid of \(1024^3\) cells and consider smoothed versions of this CIC \mbox{(over-)}density field, $\delta_{R_s}$, by multiplying it with the Fourier transform of the spherical top-hat filter of radius $R_s$ in Fourier space.
From this smoothed density field we compute the third and fourth cumulant statistics, also called skewness $S_3$ and kurtosis $S_4$, and defined by \cite{Bernardeau:2002}, e.g., as
\begin{equation}
		S_3(R_s) = \text{\LARGE $\sfrac{\text{\normalsize $\left\langle \delta_{R_s}^3 \right\rangle_{{\rm c}}$}}{\text{\normalsize $\left\langle \delta_{R_s}^2 \right\rangle^2$}}$}\,, \quad S_4(R_s) = \text{\LARGE $\sfrac{\text{\normalsize $\left\langle \delta_{R_s}^4 \right\rangle_{{\rm c}}$}}{\text{\normalsize $\left\langle \delta_{R_s}^2 \right\rangle^3$}}$} \,,\label{eq:cumulant}
\end{equation}
where the angle brackets indicate volume averages, and the connected part of the third and fourth moments are defined by:
\begin{equation}
		\left\langle \delta_{R_s}^3 \right\rangle_{{\rm c}} =
		\left\langle \delta_{R_s}^3 \right\rangle
		- 3\left\langle \delta_{R_s}^2 \right\rangle\ \left\langle \delta_{R_s} \right\rangle
		+ 2 \left\langle \delta_{R_s} \right\rangle^3\,,
\end{equation}
\begin{align}
	\left\langle \delta_{R_s}^4 \right\rangle_{{\rm c}} = &
	\left\langle \delta_{R_s}^4 \right\rangle
	- 4\left\langle \delta_{R_s}^3 \right\rangle \left\langle \delta_{R_s} \right\rangle   \nonumber \\
	                                                      & - 3\left\langle \delta_{R_s}^2 \right\rangle^2\
	+ 12 \left\langle \delta_{R_s}^2 \right\rangle \left\langle \delta_{R_s} \right\rangle^2
	- 6 \left\langle \delta_{R_s} \right\rangle^4\,.
\end{align}

Since we only use one realization for each simulation, a bootstrapping method was used to estimate the mean and variance of these quantities. Specifically, from the density fields \(\delta_{{R_s},\mathrm{num}}\) and \(\delta_{{R_s},\mathrm{den}}\) appearing in the numerator and the denominator of the ratios in eq.~(\ref{eq:cumulant}), we created an ensemble of new fields \(\delta_{{R_s},\mathrm{num}}^{i}\) and \(\delta_{{R_s},\mathrm{den}}^{i}\) with the same number of elements as the original. These elements were chosen randomly with replacement, meaning that a particular element on the original sample might not be present in the resampling, and that other elements can appear more than once. For each \(i\), the random resampling employed the same pseudo-random generator seed for the numerator and denominator of the ratio. The ratios are then estimated using the means of the one point statistics on this ensemble of resamplings. We used an ensemble of 20 resamplings.

\subsubsection{Power spectra and bispectra}\label{sec:spectra.calculation}

We define the power spectrum $P(k)$ as
\begin{equation}
	\left\langle \delta(\vect{k}) \, \delta(\vect{k}') \right\rangle = (2\uppi)^3\,\,
	\delta_{\rm D}^{(3)} \left( \vect{k} + \vect{k}' \right) \,P\left(k\right),
\end{equation}
where $k=\|\vect{k}\|$ and $\delta_{\rm D}^{(3)}$ is the Dirac delta, itself defined by
\begin{equation}
	\label{eq:dirac.fourier.transform}
	\delta_{\mathrm{D}}^{(3)} \left( \vect{x} \right) = \int \!\frac{\dd^3 k}{(2\uppi)^3} {\rm e}^{-{\rm i} \vect{k} \cdot  \vect{x}}.
\end{equation}
The power spectra are computed on-the-fly during the simulation based on an FFT of the PM grid of $2048^3$ cells with CIC assigned particles. To reduce the effect of the mass assignment, we then divide our measurements by the Fourier transform of a real-space CIC kernel \citep{Jing:2005}. We have compared our results with a measurement performed after folding the density field 64 times in each direction. This folding procedure allows to shift the range of sampled wave numbers to smaller scales \citep[see][who first proposed this method]{Jenkins:1998}. From this, we verified that our power spectra results are not influenced by the FFT grid resolution at any scale considered here.

Analogously, the bispectrum $B(k_1, k_2, k_3 )$ is defined by
\begin{equation}
	\left\langle \delta ( \vect{k}_1) \,\delta (\vect{k}_2 )\,
	\delta ( \vect{k}_3 ) \right\rangle_{\rm c}
	= (2\uppi)^3 \delta_{\rm D}^{(3)}( \vect{k}_1 + \vect{k}_2 + \vect{k}_3 ) \,B( k_1, k_2, k_3 ).
	\label{eq:bispectrum}
\end{equation}
We use the {\sc Python} package {\sc BSkit}\footnote{available from \url{https://github.com/sjforeman/bskit}} \citep{Foreman:2020} to compute the bispectra presented in this paper. This software package is based on the {\sc Nbodykit}\footnote{available from \url{https://github.com/bccp/nbodykit}} toolkit of \cite{Hand:2018} and implements a parallel version of the ``Scoccimarro estimator'' \citep[cf.][]{Scoccimarro:2000,Sefusatti:2016,Tomlinson:2019}.

We recall here the basic ideas behind this bispectrum estimator. The brute-force method integrates the triple-product $\delta(\vect{q}_1) \,\delta(\vect{q}_2)\, \delta(\vect{q}_3)$ over spherical shells of width \(\Delta k\) around \(k_1\), \(k_2\) and \(k_3\):
\begin{multline}
	\label{eq:bispec.brute.force}
	\hat{B}(k_1, k_2, k_3) = \frac{1}{(2\uppi)^3 V_{\Delta}} \int_{k_1} \dd^3 q_1 \int_{k_2} \dd^3 q_2 \int_{k_3} \dd^3 q_3 \\ \times \delta(\vect{q}_1)\, \delta(\vect{q}_2)\, \delta(\vect{q}_3)\, \delta_{\mathrm{D}}^{(3)} (\vect{q}_{1} + \vect{q}_{2} +\vect{q}_{3})\,,
\end{multline}
where
\begin{equation}
	\label{eq:bispec.triangle.density}
	V_{\Delta} = \int_{k_1} \dd^3 q_1 \int_{k_2} \dd^3 q_2 \int_{k_3} \dd^3 q_3 \,\delta_{\mathrm{D}}^{(3)} (\vect{q}_{1} + \vect{q}_{2} +\vect{q}_{3})\,,
\end{equation}
where the short-hand notation $\int_{k_i} \dd^3 q_i$ denotes integration while applying a filter with a window function  \(W(\vect{q}_i)\) that is 1 inside the spherical shell around \(\vect{k}\) and 0 elsewhere.
Inserting Eq.\,\eqref{eq:dirac.fourier.transform} in~\eqref{eq:bispec.brute.force} and using Fubini's theorem to reorder the terms, we get
\begin{equation}\label{eq:bispec.brute.force.reordered}
	\hat{B}(k_1, k_2, k_3) = \frac{1}{(2\uppi)^3 V_{\Delta}} \int \frac{\dd^3 x}{(2\uppi)^3} I_{k_1}(x)\, I_{k_2}(x)\, I_{k_3}(x)\,,
\end{equation}
where
\begin{equation}
	\label{eq:integral.delta.q}
	I_{k_j}(x) = \int \dd^3 q_j \,W_j(\vect{q_j})\,\delta(\vect{q}_j) \,{\rm e}^{-{\rm i} \vect{x} \cdot \vect{q}_j}\,.
\end{equation}
Written as in Eq.\,\eqref{eq:bispec.brute.force.reordered}, the bispectrum calculation involves three inverse Fourier transforms, their product and an integration over the real spatial domain. While the complexity is the same as a brute-force averaging over every triangle, the calculations are much more optimizable using cache efficient and parallel computations, and exploit advantages of FFT algorithms. The volume factor~\eqref{eq:bispec.triangle.density} is calculated using the same method, with a density field of value 1, and can be precomputed and reused if multiple bispectra estimations are done with the same binning.

\subsubsection{Halo finding and mass functions}
We use an inlined, on-the-fly version of the {\sc Subfind} algorithm \citep{Springel:2001} to identify gravitationally bound structures in the simulations. The underlying Friend-of-Friends (FoF) haloes \citep{Davis:1985}, defined with a linking length of $b=0.2$, must contain a minimum number of 32 particles. When we discuss mass function results, we present results for these FoF haloes. We did not find a significant difference in our results if instead spherical overdensity masses obtained at 200 times the critical density were used, once the FoF masses are corrected for discreteness effects (see discussion in \autoref{sec:massfunc_discreteness}).
To compute the mass function, we use 20 mass bins with uniform logarithmic spacing in the range \( \left[32;\ 5 \times 10^5\right]\times{\rm m_{part}^{sc}}\), where \(\rm m_{part}^{sc}\) is the particle mass in the runs with initial SC lattice. The ratios and error bars are estimated using bootstrap resampling of the catalogue of halo masses, similar to the method described in \autoref{subsubsec:density.one.point.stats}.

%%%%%%%%%%%%%%%%%%%%%%%%%%%%%%%%%%%%%%%%%%%%%%%%%%
\section{Results}\label{sec:results}

In this section, we present our study of the impact of the starting time $\zstart$ and of the order of the Lagrangian perturbation theory used to initialize the simulations. We consider various summary statistics that are commonly extracted from cosmological \(N\)-body simulations: one-point cumulants, power spectra and bispectra of the matter density field, and halo mass functions. To assess the impact of discreteness errors, we compare each case to the results from a simulation that oversamples the same initial perturbation field with four times more particles.

\begin{figure}
	\centering
	\includegraphics[type=pdf, ext=.pdf, read=.pdf, width=1.0\columnwidth]{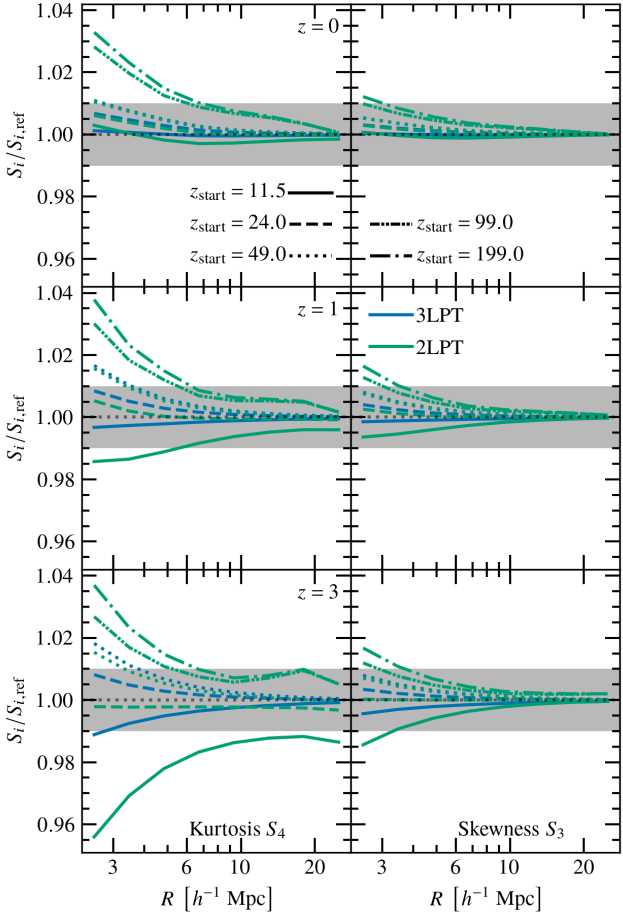}
	\caption{Dependence of one-point density field statistics on LPT order and starting redshifts. The panels show the ratios of the skewness \(S_3\) (right column) and kurtosis \(S_4\) (left column) as a function of smoothing scale $R_s$ of simulations started at different redshifts (line styles), with different orders of LPT (line colors), with respect to the FCC run started at \(\zstart = 24\) and 3LPT initial conditions. We find these to be well converged in all cases, except the kurtosis when 2LPT is  initialized too late, while 3LPT still yields about one per cent accuracy on $S_3$ and $S_4$ even when  initialized as late as $\zstart=11.5$. }
	\label{fig:one.point.statistics}
\end{figure}

\subsection{One-point statistics}\label{subsec:one.point.statistics}

We start by considering cumulants, $S_n$, of the matter density field. The non-linear gravitational collapse introduces non-Gaussianity in the originally purely Gaussian density field, which can be captured in these cumulants. In addition, high-order cumulants of the matter density field are sensitive to the order of the LPT used to set up the simulation and the starting redshift \citep[e.g.][]{Crocce:2006}.

Our measurements of the skewness and kurtosis of the density distribution function of the  matter density field smoothed on different scales $R_s$ are presented in \autoref{fig:one.point.statistics}. Specifically, we show ratios  of \(S_3\left(R_s\right)\) (left column), and \(S_4\left(R_s\right)\) (right column) at redshifts $z=0$, 1 and 3 (first to third row)  initialized at different times (represented by the line styles) and with different LPT orders (represented by the line colours) with respect to the 3LPT reference run  initialized with an oversampled FCC lattice at \(\zstart = 24\).

For a large enough smoothing radius, the filtered density field approaches a Gaussian field, and the agreement between simulations should be perfect. This is what we see indeed for \(S_3\) with both 2 and 3LPT. For \(S_4\) however, while 3LPT converges correctly, 2LPT does not agree with our reference simulation even on large scales $R_s\gtrsim10\,h^{-1}{\rm Mpc}$. At the latest starting time we consider, \(\zstart = 11.5\), the disagreement for 2LPT is larger than \(1\%\) at \(z = 3\), reducing to \(0.2\%\) at \(z = 0\) for \(R_s = 25h^{-1}\ \mathrm{Mpc}\).
For \(\zstart = 24\) the difference stays within \(1\%\) at all scales for \(z \leq 3\) but the transients are still visible in \(S_4\) with 2LPT. \cite{Munshi:1994} have shown in the case of  spherical collapse that \(n\)LPT reproduces the one-point statistics up to \(S_{n+1}\). We confirm here that indeed 2LPT correctly predicts \(S_3\) but not \(S_4\) in the large-$R_s$ limit, while 3LPT is correct for both. Likewise, we expect all cumulants higher than 4 to be incorrect even with 3LPT. A similar result for 2LPT was presented by \cite{Crocce:2006}.

The deviations from Gaussianity that are produced by the gravitational collapse of over-densities affect the smallest scales first. The differences with respect to the reference simulation in \autoref{fig:one.point.statistics} can be interpreted as the presence of small-scale transients due to the truncation of the LPT series. For the higher starting redshift (\(\zstart = 24\)), they are all under \(1\%\) even for the smallest shown scale of \(2.5h^{-1}\ \mathrm{Mpc}\) (since the mean particle separation is $\sim 1h^{-1}\ \mathrm{Mpc}$, we did not probe smaller scales). For the lower starting redshift \(\zstart = 11.5\), the disagreement with 2LPT for \(S_3\) is under \(2\%\) at \(z = 3\) and goes under \(1\%\) for \(z \leq 1\). For \(S_4\) the relative difference is \(>3\%\) at \(z = 3\), \(2\%\) at \(z = 1\) and under \(0.5\%\) at \(z = 0\). This is roughly consistent with the earlier results of \cite{Tatekawa:2007} for 3LPT including only longitudinal modes, who however were not studying per cent level agreements, so that a detailed comparison is not possible.

Of course, the lower starting redshift ($z_{\rm start}=11.5$) is extremely late by traditional wisdom of when to start a simulation, and based on our analysis of the convergence radius of LPT in \autoref{sec:convergence_radius}, it is also dangerously close to the time when we expect LPT to break down (for the considered resolutions). It is thus no surprise that 2LPT does not fare well in this case.  In contrast, however, 3LPT performs at the per cent accuracy level even for such extremely late starts. For example, in the case of \(S_4\), the difference is already under \(2\%\) at \(z = 3\), and  for \(S_3\) it is always below \(1\%\).

%%%%%%%%%%%%%%%%%%%%%%%%%%%%%%%%%%%%%%%%%%%%%%%%%%%%%%%%%%%%
\subsection{Two-point statistics}\label{subsec:two.point.statistics}

We next investigate the impact of the initial conditions and starting time on the late-time two-point statistics, as quantified by the matter density power spectrum. First we quantify the magnitude of discreteness errors as a function of starting time, then the impact of LPT order and starting time on the accuracy of matter power spectra. We will consider a one per cent agreement with a reference solution to the largest possible wave number $k$ as the benchmark of accuracy for all spectra.

\subsubsection{Discreteness errors on power spectra}\label{sec:discreteness_power}

\begin{figure}
	\centering
	\includegraphics[clip,type=pdf, ext=.pdf, read=.pdf, width=0.98\columnwidth]{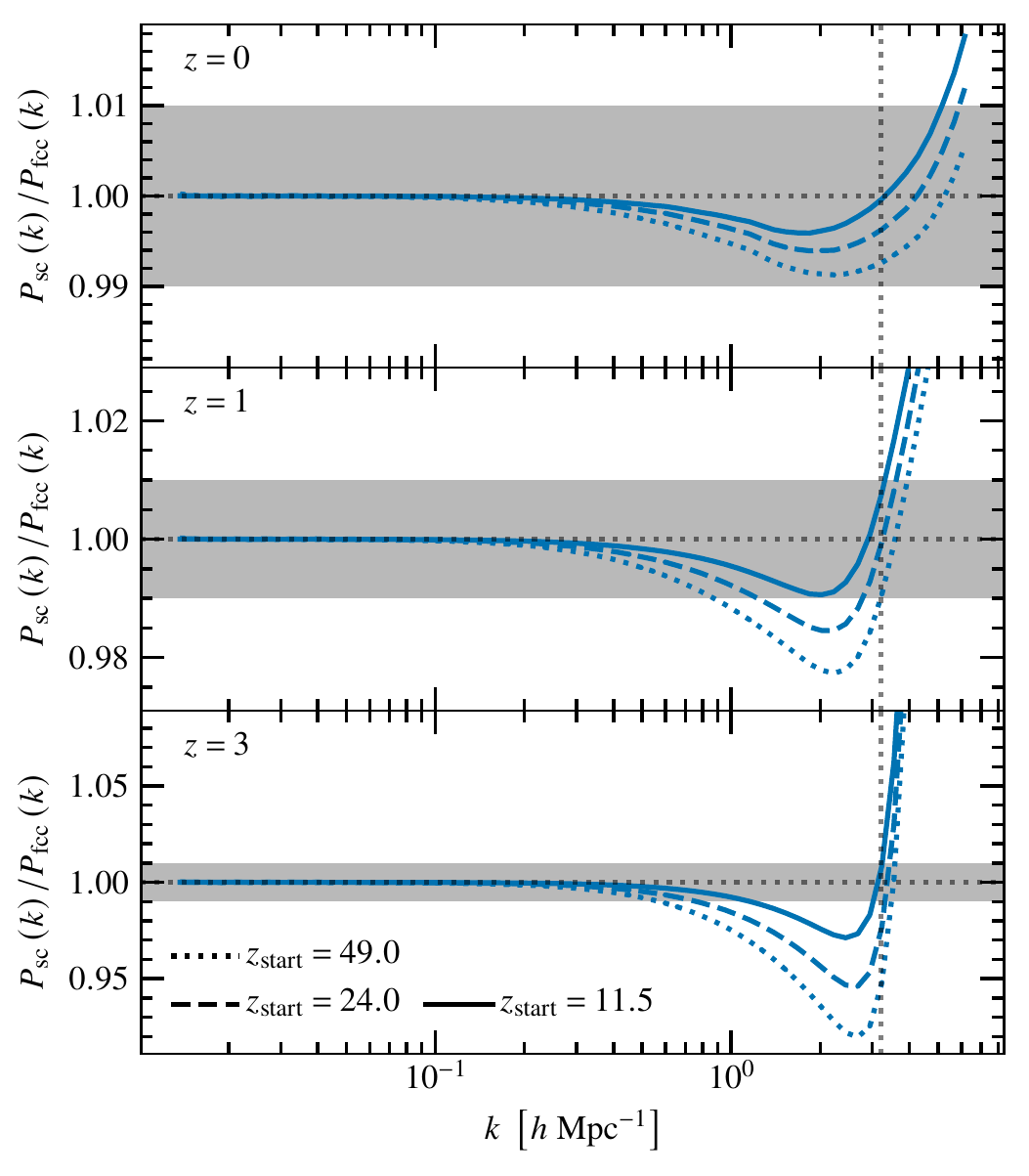}
	\caption{Impact of particle discreteness on the matter power spectrum evolution: Ratios between the power spectra of the SC and oversampled FCC runs for different starting reshifts (different line styles) at $z=0$, $1$ and $3$ (top to bottom panels). All simulations shown use 3LPT ICs, but the results do not depend on the LPT order. The grey shaded area indicates per-cent-level agreement with the reference. The vertical dashed line indicates the particle Nyquist wave number of the initial perturbation field. Discreteness effects lead to a time-dependent suppression close to the particle Nyquist wave number. }
	\label{fig:powerspec.discreteness}
\end{figure}

In order to investigate the impact of particle discreteness on the power spectrum, we ran the same initial perturbation spectrum with particles initially placed on an SC lattice, as well as on a 4 times oversampled FCC lattice (see our discussion in \autoref{sec:lattice_PLT} for more details). The initial conditions were generated using 3LPT at starting redshifts $z_{\rm start}=49$, $24$ and $11.5$. In \autoref{fig:powerspec.discreteness}, we show the ratio of the SC to the FCC runs at redshifts $z=3$, $1$ and $0$. At $z=0$, all power spectra agree to better than one per cent up to the particle Nyquist wave number, irrespective of the starting time. However, at higher redshifts, the SC lattice shows an important suppression of the small scale structures near the particle Nyquist wave number due to discreteness errors. At $z = 3$ for $\zstart = 49$, the underestimation of the SC compared to FCC is about $8\%$, down to $\sim3\%$ at $z = 1$. For the late start \(\zstart = 11.5\), we find a maximum suppression of $\sim2\%$, $\sim1\%$ and $\sim0.4\%$ at $z = 3$, 1 and 0, respectively. As a consequence, the power spectra agree to better than one per cent for $k\gtrsim k_{\rm Ny}/3$ at $z=3$ and for all $k\gtrsim k_{\rm Ny}$ at $z=1$ and 0. While not shown in the figure, we found that for a fixed starting redshift, the suppression does not depend on the order of LPT used. The independence from the LPT order clearly indicates an origin in discreteness errors, and the shape of the power suppression is indeed very similar to the scale-dependent PLT growth factors \citep[cf.\ Fig.~1 of][]{Joyce:2005} indicating that the fluid ICs relax to the discrete evolution in the quasi linear regime, and increasingly so, the earlier the starting time.

We thus find that {\sl the higher the starting redshift is, the greater is the suppression of the high \(k\) modes, irrespective of the order of LPT.} The effect peaks at intermediate redshifts, before the high-$k$ part of the resolved power spectrum becomes fully dominated by collapsed haloes. A similar suppression of high-$k$ power that improves the better the non-linear scale is resolved has also been reported by \cite{Schneider:2016}, who also investigated the additional dependence on the gravity solver employed in the simulations. We leave an assessment of whether/how much our results depend on the choice of \(N\)-body gravity solvers for future work.

\subsubsection{Dependence of power spectra on LPT order \& starting time}\label{sec:ICs_power}

\begin{figure}
	\centering
	\includegraphics[clip,type=pdf, ext=.pdf, read=.pdf, width=0.98\columnwidth]{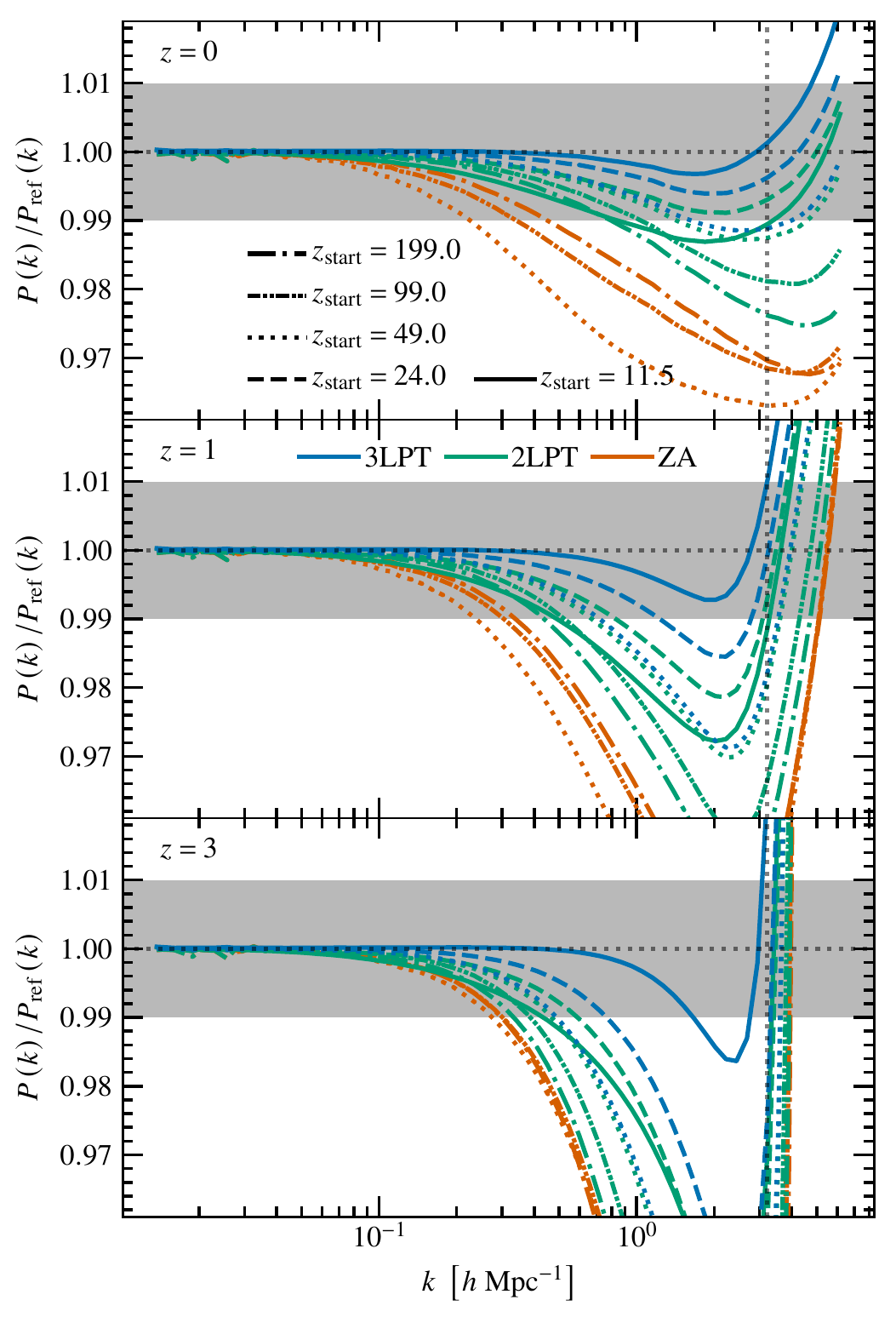}
	\caption{Impact of starting time and LPT order on the matter power spectrum evolution: Ratios of power spectra for different orders of LPT (line colours) and starting redshifts (line styles) at $z=0$, $1$ and $3$ (top to bottom panels), w.r.t. the FCC 3LPT $z_{\rm start}=24$ reference simulation. The grey shaded area indicates per-cent-level agreement with the reference, and the dashed vertical line indicates the particle Nyquist wave number. Higher order LPT and later starting times improve agreement with the reference run.}
	\label{fig:powerspec.ratios}
\end{figure}

\autoref{fig:powerspec.ratios} shows ratios of power spectra at redshifts $z = 0$, 1 and 3 (panels top to bottom) of simulations starting from different initial conditions with respect to a reference simulation. Specifically, we vary the order of the Lagrangian perturbation theory from the ZA (orange) to 2LPT (green) to 3LPT (blue lines), and the starting time between $z_{\rm start}=199$, $99$, $49$, $24$ and $11.5$ (as indicated by the different line styles). Note that we have not run all combinations of LPT order and $z_{\rm start}$. We use a simulation  initialized with 3LPT at $z_{\rm start}=24$ and using an FCC lattice (i.e. four times more particles) as our reference simulation. This reference solution should suffer from much reduced discreteness effects due to the significantly higher particle number and increased symmetry of the lattice.

The results shown in \autoref{fig:powerspec.ratios} indicate clearly that for a simulation with our resolution, 3LPT performs best at $z=3$ and is accurate for $k\lesssim k_{\rm Ny}/2$ if started at $z_{\rm start}=11.5$, while performing increasingly similar to 2LPT for earlier starts, with both being accurate roughly for $k\lesssim k_{\rm Ny}/4$ and  $\lesssim k_{\rm Ny}/5$ when started at $z=24$ and $z=49$, respectively. Note that 2LPT, in stark contrast to 3LPT, performs very poorly for the latest start $z_{\rm start}=11.5$, which just reflects that this is indeed a too late start for  second order LPT. The ZA runs, for which we only considered much earlier starting times $z_{\rm start}=49$, $99$ and $199$, are all accurate only to $\sim k_{\rm Ny}/10$, and have the largest errors overall.

As structures collapse into haloes over time, the power spectra become significantly less sensitive to the initial conditions as evidenced by our results at $z=1$ and $z=0$. The 3LPT run with the latest starting time $z_{\rm start}=11.5$ is now accurate up to the particle Nyquist wave numbers ($z=1$) or even beyond ($z=0$). This is also true for 2LPT and 3LPT at $z=0$ if started at $24$. All other runs are less accurate, to varying degrees. Most notably, and as expected from our discussion of discreteness effects, the ZA runs converge only extremely slowly with increasingly earlier starting time, and arguably converge to the discrete solution rather than the fluid solution. Even for the earliest start we considered, $z_{\rm start}=199$, the ZA run is only accurate for $k\lesssim k_{\rm Ny}/8$ ($\lesssim k_{\rm Ny}/9$)  at $z=0$ ($z=1$). We note that for early starts, of course, ZA and 2LPT are consistent at $z=0$ within about one per cent of each other (cf. the dash-dotted orange and green lines), which is what has also been found by \cite{Schneider:2016} (see their Fig.~3), and the general picture for the poor performance of ZA vs. 2LPT has already been outlined by \cite{Crocce:2006}.

For 2LPT, as for 3LPT, a later start gives the best results with $z_{\rm start}=24$ among the runs we considered, but the latest possible start is of course earlier for 2LPT than for 3LPT. This is in fact perfectly consistent with the analysis for 2LPT of \cite{Nishimichi:2019} who followed a different approach to find the optimal starting redshift, by finding that $z_{\rm start}$ for which the small-scale power spectrum has the highest amplitude, thereby finding the ``sweet spot'' between particle discreteness and LPT truncation transients. Their $250\,h^{-1}{\rm Mpc}$ box with $256^3$ particles has the same Nyquist mode as our runs, and they find the optimal start to be around $z_{\rm start}\simeq30$. For comparison, we note that the so-called Euclid Flagship simulation \citep{Potter:2017} and those used in the EuclidEmulator \citep{Knabenhans:2019} were initialized at $z=200$ with ZA. Our results indicate that their results could be systematically biased low by about 4\% at $k \sim 4\,h{\rm Mpc^{-1}}$. In addition, the simulations of \citep{Angulo:2020}, were initialized at $z=49$ with 2LPT, which should be biased low by 1.5\% on the same scales.

Overall, we thus find that a picture emerges that fits nicely with our analysis of the convergence radius of LPT presented in \autoref{sec:conv_numerical}. The best results are obtained with the latest start at the highest order of LPT. These results should be compared against the ones displayed in~\autoref{fig:convergence_sigma_amax}. The starting redshifts of 49, 24 and 11.5 are placed respectively just below the 99.7th, just above the 95th, and just above the 68th percentile of the convergence radius plot for the box size and resolution considered here (cf. symbols in \autoref{fig:convergence_sigma_amax}). It thus appears that for the best accuracy on the power spectrum, pushing the starting time as low as the 68th percentile yields the best results. We did not consider even later starting times, but they might yield a small further improvement for 3LPT. One has to caution against pushing this too far however, as beyond the radius of convergence, higher orders of LPT introduce higher errors.

\begin{figure}
	\centering
	\includegraphics[type=pdf, ext=.pdf, read=.pdf, width=0.98\columnwidth]{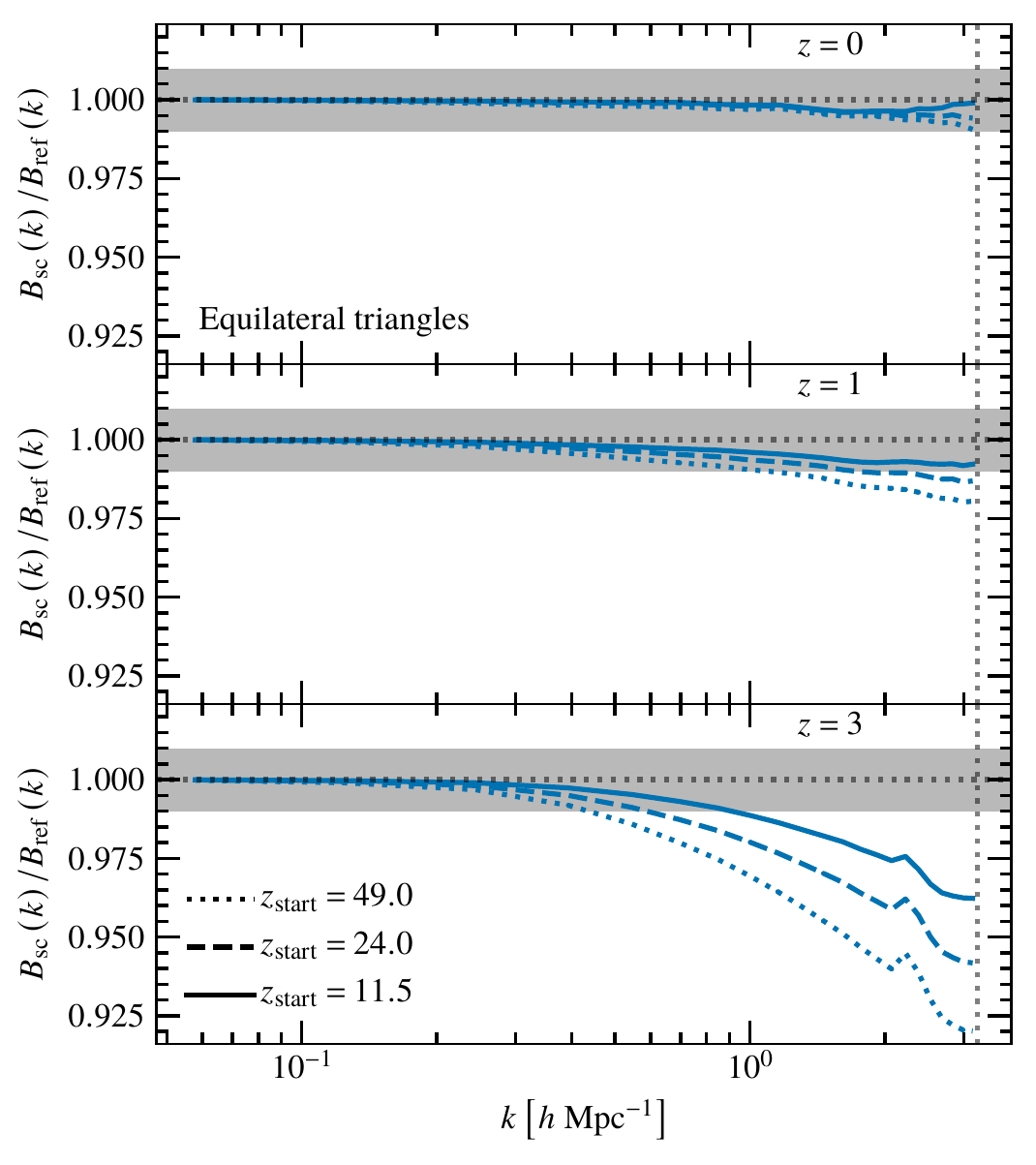}
	\caption{Same as \autoref{fig:powerspec.discreteness} but showing the impact of particle discreteness on the equilateral bispectrum $B(k) \equiv B(k,k,k)$.}
	\label{fig:bispec.discreteness}
\end{figure}

\begin{figure}
	\centering
	\includegraphics[type=pdf, ext=.pdf, read=.pdf, width=0.98\columnwidth]{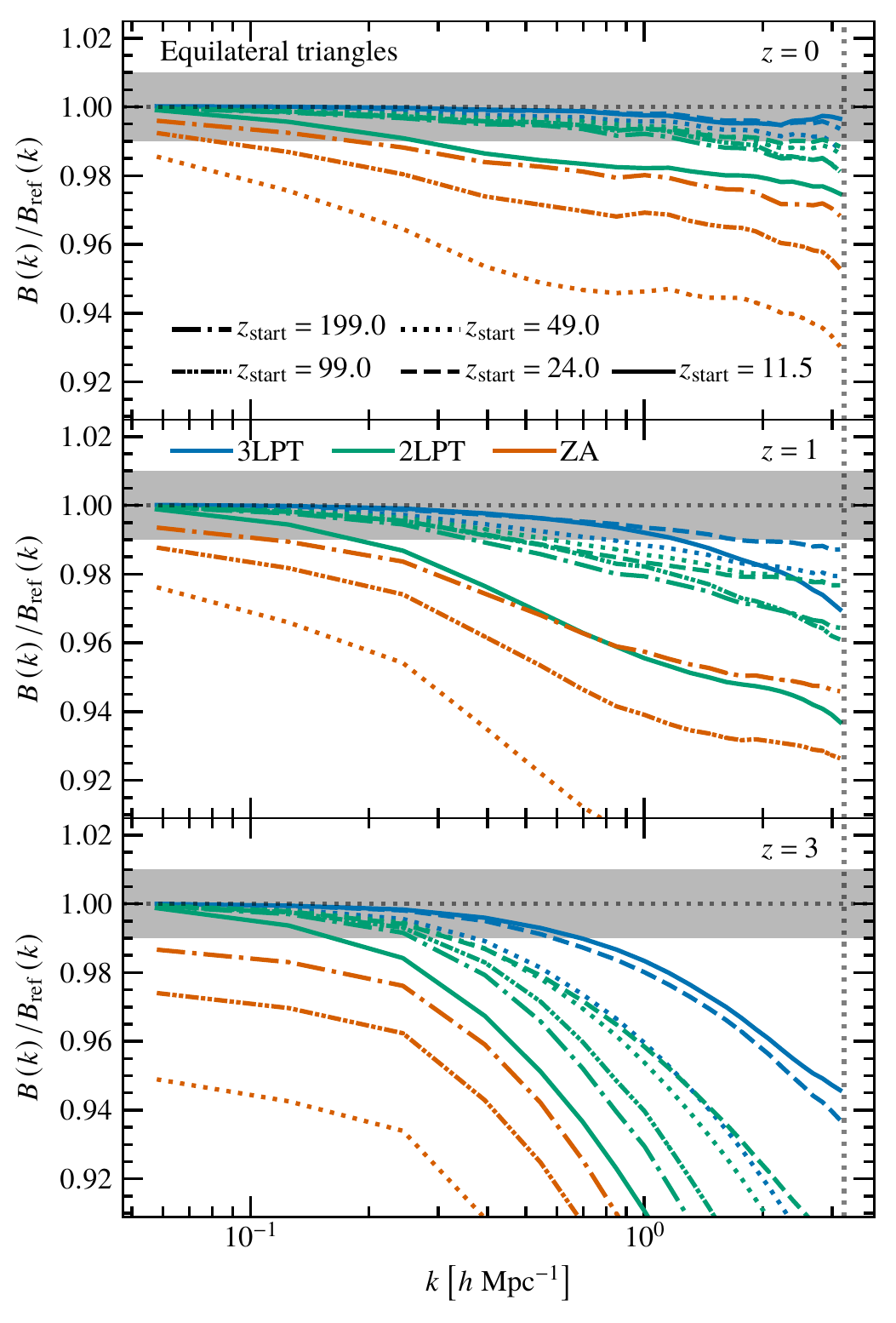}
	\caption{Same as \autoref{fig:powerspec.ratios} but showing the impact of starting time and LPT order on the equilateral bispectrum. The impact of 3LPT on the bispectrum is more dramatic: At $z=1$ we were not able to achieve per cent level agreement at this resolution with 2LPT to small scales, and ZA is so off that it should not be used for precision studies. 3LPT allows sub-per cent errors on the bispectrum at $z=0$ up to the Nyquist mode.}
	\label{fig:bispec.equilateral}
\end{figure}

\subsection{Three-point statistics}\label{subsec:three.point.statistics}

In this section, we repeat the analysis of the previous section for bispectra. In order to simplify the analysis, we exclusively focus on equilateral bispectra $k=k_1=k_2=k_3$, which are one-dimensional in that they depend only on a single scalar $k$, and not on the direction~$\vect{k}$ (due to the assumed statistical isotropy), and thus allow us to probe the scale-dependence up to the particle Nyquist wave number most conveniently. We analyse the bispectra for the same simulations as those in \autoref{subsec:two.point.statistics}. Again, we first quantify the magnitude of discreteness errors as a function of starting time, then the impact of LPT order and starting time on the accuracy of the bispectra. As before, we will consider a one per cent agreement with a reference solution in order to quantify accuracy.

\subsubsection{Discreteness errors on bispectra}
We again quantify the impact of particle discreteness on bispectra by taking ratios of bispectra measured from simulations that sample the same perturbations using an SC and a four times oversampled FCC lattice (cf. \autoref{sec:lattice_PLT}).
As before, we vary the starting redshift of the 3LPT initial conditions between \(z_{\rm start}=49\), \(24\) and \(11.5\). The results of this study are shown in \autoref{fig:bispec.discreteness}, and are in broad agreement with our previous analysis of discreteness effects on the power spectrum. Again, we see a time and scale dependent suppression which is the strongest close to the particle Nyquist wave number. For the latest starting time, \(z_{\rm start}=11.5\), agreement with the reference run is at the sub-per cent
level at \(z=0\) and \(1\) up to \(k_{\rm Ny}\), with increasing errors for the earlier starts. At \(z=3\) errors are significantly larger, but still with later starts faring better. These results are in tension with the interpretation given by \cite{McCullagh:2015} who report a convergence of bispectra only for early enough starting times. We will discuss this more below.

\subsubsection{Dependence of bispectra on LPT order \& starting time}
We next investigate the dependence of the accuracy of bispectra on the initial conditions varying LPT order and starting time. \autoref{fig:bispec.equilateral} shows the ratios of equilateral bispectra w.r.t. the reference run (3LPT, oversampled FCC lattice, $z_{\rm start}=24$). We vary the orders of LPT (line colour) and starting redshift (line style), and present measurements at \(z = 0\), 1 and 3. One notes immediately that for bispectra, the impact of the order of LPT used to set up the simulations is much more pronounced than for the power spectra presented in the previous \autoref{sec:ICs_power}. At all times and all scales, and irrespective of the starting time, the bispectrum obtained from ZA initial conditions disagrees with the reference run, with differences dramatically increasing at earlier times. Changing to 2LPT ICs improves the situation dramatically, as has been reported before \citep[e.g.][]{McCullagh:2015,Baldauf:2015}. However, even for 2LPT, the bispectra agree at $z=1$ with the reference solution only for $k\lesssim k_{\rm Ny}/15$ (if we disregard the latest starting time $z_{\rm start}=11.5$). The dependence on starting time is however non-trivial for 2LPT: there is clearly a sweet spot for the runs starting at $z\approx24-49$, with worse agreement with the reference run for both earlier and later starts. While disagreeing with the reference run, the results appear to converge for increasingly earlier starting times ($z_{\rm start}=99$ and $199$), which are however strongly suppressed w.r.t. the reference run at $k\sim 1 h{\rm Mpc}^{-1}$. Generally, the situation improves with the growth of non-linear structure on increasingly larger scales at $z=0$. Now the later starting times $z_{\rm start}=49$ and $24$ agree with the reference run at $k\lesssim k_{\rm Ny}/4$, while the early starts agree perfectly with one another, but not with the reference run for $k\gtrsim k_{\rm Ny}/8$. We are arguably witnessing the convergence to the discrete solution with the increasingly higher starting times here. The best results are obtained with 3LPT and again the latest possible starts, which agree with the reference at $z=0$ at all scales we investigated. At $z=1$, the agreement is still excellent for the $z_{\rm start}=24$ run ($k\lesssim k_{\rm Ny}/2$). It is slightly less good for the latest start ($z_{\rm start}=11.5$), indicating that this is a slightly too late start even for 3LPT. At $z=3$ the 3LPT runs are accurate only for $k\lesssim k_{\rm Ny}/15$ but have overall the smallest errors.

In contrast to the results of \cite{McCullagh:2015}, who suggest using very early starting times, our results appear to indicate quite the opposite. Since bispectra are impacted by particle discreteness just like power spectra, discreteness errors leading to a deviation from the fluid limit accumulate more, the higher the starting redshift and the smaller the perturbations.

In summary, we find that, as for the power spectra, the bispectrum accuracy is best when high-order LPT is combined with a very late start. \(N\)-body simulations starting from 3LPT at our relatively low resolution certainly can be accurate to predict the bispectrum at $z=0$ to better than one per cent up the particle Nyquist wave number.

Note that there have been concerns in the literature about the accuracy with which \(N\)-body simulations predict the matter bispectrum \citep[e.g.]{Schmittfull:2013,Baldauf:2015,Hung:2019}. For instance, \cite{Baldauf:2015} assumes a systematic error in their simulated results to account for lack of convergence when ZA or 2LPT is used. In contrast, our results indicate that once discreteness and truncation are taken into account, \(N\)-body simulations can provide very reliable and accurate results.

%%%%%%%%%%%%%%%%%%%%%%%%%%%%%%%%%%%%%%%%%%%%%%%%%%%%%%%%%%%%
\subsection{Halo mass functions}\label{subsec:halo.mass.functions}

\begin{figure}
	\centering
	\includegraphics[type=pdf, ext=.pdf, read=.pdf, width=0.9\columnwidth]{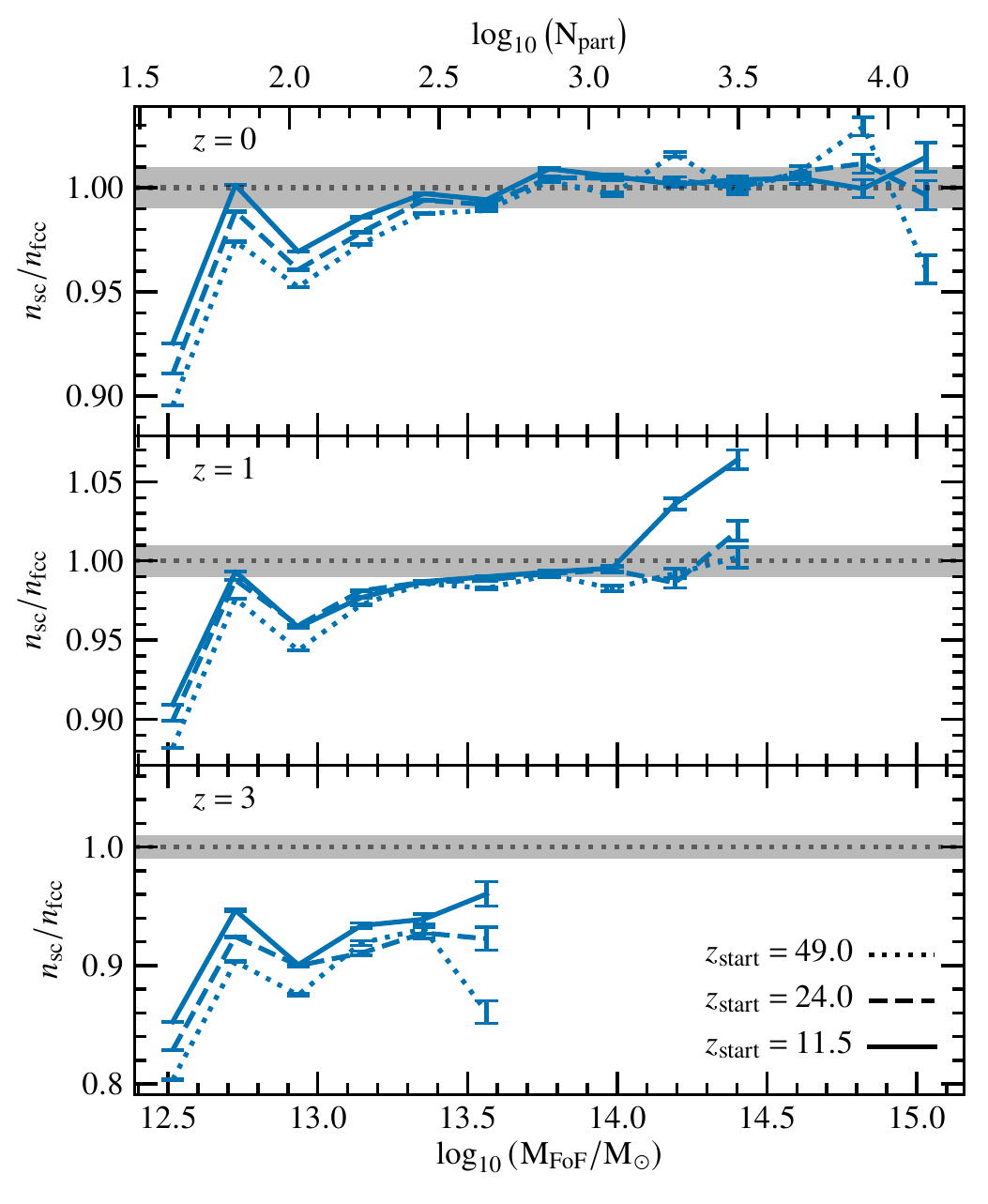}
	\caption{Same as \autoref{fig:powerspec.discreteness} but for the halo mass function. We indicate halo mass in the bottom $x$-axis, and corresponding particle number in the SC run in the top $x$-axis. The FoF-halo masses have been corrected using the Warren et al. (2006) correction (see text for details).}
	\label{fig:mass.function.discreteness}
\end{figure}

\begin{figure}
	\centering
	\includegraphics[type=pdf, ext=.pdf, read=.pdf, width=\columnwidth]{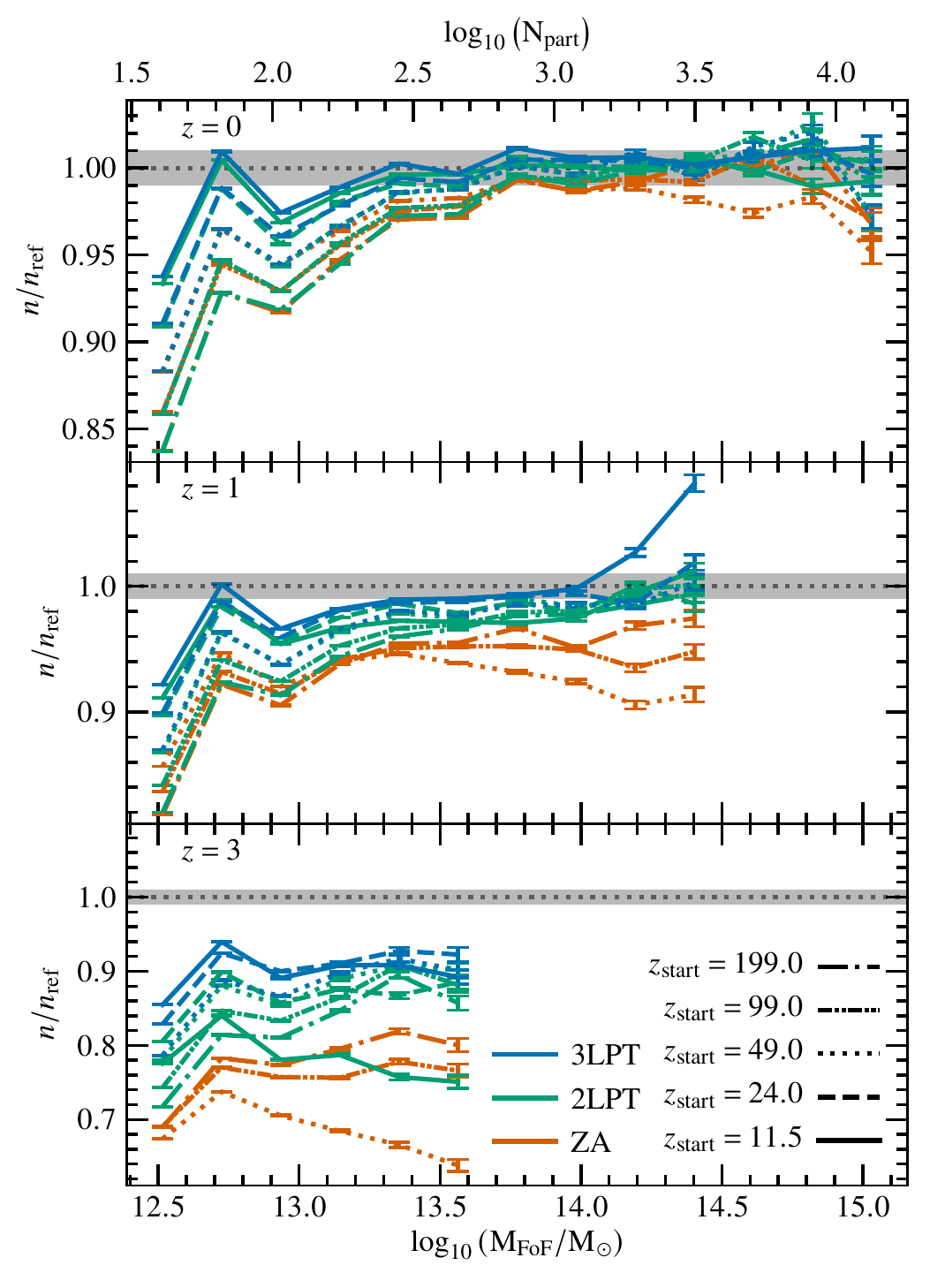}
	\caption{Same as \autoref{fig:powerspec.ratios} but for the halo mass function, showing a strong dependence on starting time and LPT order, especially at high redshift. }
	\label{fig:massfunction.ratio}
\end{figure}

Finally, we also consider the abundance of collapsed structures -- as quantified by the halo mass function -- as a benchmark for the dependence of simulation results on the initial conditions. To this end, we essentially repeat once again the analysis performed already in the sections before. We first quantify the impact of discreteness on the mass function, and in a second step the dependence on starting time and order of LPT on the abundance of haloes.

\subsubsection{Discreteness errors in the mass function}\label{sec:massfunc_discreteness}

In \autoref{fig:mass.function.discreteness}, we show the ratio of mass functions obtained from 3LPT IC runs (starting from an SC lattice) with varying starting time to our usual reference run (3LPT, oversampled FCC lattice, $z_{\rm start}=24$). The FCC reference has four times more particles, meaning that haloes of a given mass are much better resolved in that run, but does not introduce further small-scale modes (see discussion in \autoref{sec:lattice_PLT}). In this sense, our convergence test is different from the usual convergence tests of halo mass functions \citep[e.g.][]{Jenkins:2001,Tinker:2008}, where simply different ranges of scales of the full \(\Lambda\)CDM spectrum are resolved. In our results, we show mass functions for FoF halos. The FoF algorithm is known to be biased at low particle numbers due to percolation noise. \cite{Warren:2006} have proposed to correct the mass of a FoF group due to discreteness errors at a finite number of particles $N_{\rm part}$ as
\begin{equation}
	M_{\rm FoF} = M_{\rm FoF, raw} \left(1-N_{\rm part}^{-0.6}\right) \,,
\end{equation}

\noindent which we have applied before computing the mass functions (note that this is an ad-hoc correction which might take a different form depending on the mass resolution or even starting redshift). It is also entirely possible that an SC and an FCC lattice lead to somewhat different discreteness errors in FoF haloes, which we have not investigated but do not believe to strongly impact our conclusions.

After applying this correction, our ratio plots are very similar to those  obtained when using spherical overdensity haloes (not shown) for an overdensity of 200 times the critical density,  which also underpredicts the number of haloes at the low mass end compared to the higher resolution simulation. Arguably, a more optimal correction can be performed to push up the low-mass end of the FoF-mass function \citep[cf. e.g.][]{Nishimichi:2019}, but such avenues shall not be our concern here. We also found that without the correction, the FoF mass function ratio is biased high at the per cent level at all masses due to ``Eddington bias'' \citep{Eddington:1913}, which is implicitly corrected by the ``Warren'' correction.

Looking at our simulation results, we first notice in \autoref{fig:mass.function.discreteness} a systematic drop below $\sim300$ particles at late times $z=0,1$ which is more severe for the earlier start $z_{\rm start}=49$ than the later starts. This undershooting is of the order of $\lesssim 5$ per cent for haloes of $\sim100$ particles. Finally, at the earliest time $z=3$, we see a systematic (mass-independent)  bias in the mass function that depends only weakly on the starting time of the simulation. While at this early time, the haloes in our simulations are not well resolved (all below 1000 particles), the mass function is low by 10 per cent, which is much more than at the later times. From our previous analysis of power- and bispectra, we know that simulation results are particularly sensitive to the ICs at this early time.

\subsubsection{Impact of LPT order and starting time on the mass function}\label{sec:massfunc_ics}

Finally, we show in \autoref{fig:massfunction.ratio} ratios of mass functions of dark matter haloes relative to the reference run (3LPT, oversampled FCC lattice, $z_{\rm start}=24$) for different orders of LPT and starting redshifts. In general, we find a strong dependence on both the order and starting time.

Let us first focus on the latest time, $z=0$, when structures are more developed, and we have many well resolved haloes. We find that at the high mass end, all ICs converge to the same mass function, in per cent level agreement with the reference run. At the low-mass end, we find however a very strong dependence on the starting time of the simulation, but not the LPT order used. The earliest starts are low by more than 5 per cent at the $\sim100$ particle scale. In addition, the ZA runs for the latest time we ran with ZA, $z_{\rm start}=49$ are low at 2-3 per cent even at high masses. This picture is consistent with earlier studies that also did not find a strong dependence of the late-time mass function on starting time \citep[e.g.][in the pre-precision cosmology era]{Jenkins:2001,Tinker:2008,Knebe:2009}, and \cite{Reed:2013} and \cite{Nishimichi:2019} who find per-cent level convergence for moderately early starts with 2LPT.

At the intermediate time, $z=1$, a more mixed picture appears in that the ZA runs are low by at least 5 per cent at all masses, and we see a weak additional dependence on the LPT order used, with 3LPT ever so slightly better converged than 2LPT.

Finally, at early times, $z=3$, we find a very strong dependence on the order of LPT used. ZA predicts significantly (by more than 20 per cent) fewer haloes at fixed mass than the reference run, even when started early ($z_{\rm start}=99$ and $199$). This is improved by 2LPT to about 10-15 per cent if started at $z_{\rm start}=24$ or 49. The 3LPT runs are closer to the reference, and show a weaker dependence on starting time than 2LPT, but are still $\sim10$ per cent low.

At all times, the very early starts, $z_{\rm start}=99$ and $199$, show a significantly lower abundance of low mass haloes w.r.t. the later started runs when 2LPT and 3LPT are used. At the later times $z=0$ and $1$, both 2LPT and 3LPT give results that are within about 2 per cent of one another for starting times of $z_{\rm start}=24$ or even $11.5$. The 2LPT run with $z_{\rm start}=11.5$ however is clearly off at early times, $z=3$. Again, consistent with the power- and bispectrum results, we find that discreteness effects impact most strongly the early starts, while late starts with high-order LPT give the most converged results. Only for the ZA runs did we actually observe an improvement when starting earlier.

As has been demonstrated by \cite{Reed:2013} and the recent study of \cite{Ludlow:2019}, the absolute convergence of the halo mass function at the low-mass end also depends on further variables, such as  force resolution and time-stepping. Our results should therefore ultimately be subjected to further studies that include also variations of time and force integration parameters to determine the ultimate errors on the mass function at the few-particle, low-mass end.

%%%%%%%%%%%%%%%%%%%%%%%%%%%%%%%%%%%%%%%%%%%%%%%%%%
\section{Summary and Conclusions}\label{sec:summary}

In this paper, we have presented a rigorous analysis of the interplay of discreteness effects and the order truncation in Lagrangian perturbation theory used to set up initial conditions for cosmological \(N\)-body simulations. Our findings strongly suggest, contrary to common wisdom, that in order to be most economical (i.e., the most accurate to the smallest possible scales with the least computational resources), \(N\)-body simulations should be initialized with the highest possible order of LPT at the latest possible time. In that case, we are able to achieve the, admittedly, ambitious goal of sub-percent level convergence in the matter power and bispectrum all the way to the particle Nyquist wave number. We shall summarize the arguments leading up to this conclusion next.

We have considered simulations based on initial conditions that use up to third order in Lagrangian perturbation theory (3LPT). The numerical implementation and proofs of correctness of the numerical implementation are somewhat more involved than for low-order LPT (see~\autoref{sec:numerical_implementation_lpt}). In particular, high-order LPT involves the convolution of non-linear fields and thus, numerical implementations  should be de-aliased. We demonstrate that for efficiency, at least up to 3LPT, this can however be disrespected in IC generation, since errors decay away quickly for simulations evaluated at low redshift.

The particle discretization used in an \(N\)-body simulation only approximates the underlying dark matter fluid, which manifests itself in a deviation from the evolution expected in the continuum limit. We have shown in \autoref{sec:lattice_PLT} that this results in an amplitude suppression close to the particle Nyquist wave number of the initial particle grid in both power-  and bispectra. When correcting for the linear discreteness error \citep[using particle linear theory, cf. e.g.][]{Joyce:2005}, we found that the \(N\)-body simulation agrees with 3LPT to about 1-2 per cent up to the particle Nyquist wave number at much later times than are typically used to initialize \(N\)-body simulations.  At the same time, the \(N\)-body system,  initialized as a fluid, quickly relaxes to the evolution of the discrete system \citep[cf. also][]{Garrison:2016}, particularly so while perturbations are still relatively small during the quasi-linear stage of the evolution. As a consequence, it follows that the earlier the starting time, the more closely will the evolution follow the discrete and not the fluid solution. It is thus obvious that a late starting time is preferable, since it initializes the simulation using fluid perturbations that are already relatively large. 
Subsequent nonlinear evolution, and the transfer of power from large to small scales,
will reduce the importance of initial discreteness errors, yielding progressively better
convergence.

Our conclusions are in contrast with traditional wisdom (see however \citealt{Nishimichi:2019}), that would tell us that perturbation theory is increasingly more accurate at early times, and so that cosmological \(N\)-body simulations should be initialized early. For instance, the Euclid Flagship simulation follows this guideline and starts at $z=200$.

A late start of a cosmological \(N\)-body simulation then leads to the question of how late one could  start. The answer depends, of course, on the  convergence radius of Lagrangian perturbation theory with \(\Lambda\)CDM random initial conditions, as well as on the truncation error introduced by  fixed-order LPT. As regards to the former, while theoretical predictions by \cite{Zheligovsky:2014} and \cite{Rampf:2015} revealed a finite but lower bound on the convergence radius of LPT, here we have performed provided a novel quantitative analysis that demonstrates the validity of LPT at much later times; see~\autoref{sec:ratiotest}. We found that the convergence radius follows relatively simple scaling laws with the RMS amplitude of density fluctuations at the resolution scale (see~\autoref{fig:convergence_sigma_amax}). We then assessed the impact of the LPT truncation error using standard convergence tests based on various summary statistics. In particular, we demonstrated that with 3LPT, the truncation error is so much reduced that it could be ignored even for very late starts.

In all cases, we tested convergence against a simulation with four times better mass resolution, while using the same perturbation modes. This is in contrast to previous convergence studies, that usually alter the ICs by adding more perturbations when resolution is increased. In \(\Lambda\)CDM, adding new modes implies adding more non-linearities at a given time.
This, in our reasoning, impacts the starting time and truncation errors, and thus makes any rigorous tests of LPT truncation errors and assessment of transients questionable.
For production runs, the increase in particles is likely better invested in sampling also additional modes -- which should however be investigated in future work.

After investigating in \autoref{sec:results} the dependence on starting time and LPT truncation of one-point statistics of the density field,
density power- and bispectra, as well as halo mass functions, we can summarize our main findings as follows.
\begin{enumerate}
	\item 3LPT initial conditions with the latest starting redshift we considered, namely $z_{\rm start}=11.5$, generally yielded the most accurate results in all summary statistics w.r.t.\ our better resolved reference simulation. Furthermore, memory-intensive de-aliasing procedures may be safely ignored as the evolved aliasing errors are of order per mille for all considered observables for $z\lesssim3$. Note however that this optimal starting time is for the mass resolution we consider here, and increases at higher mass resolution (cf. \autoref{fig:convergence_sigma_amax} for the scaling).
	\item In contrast, the ZA is almost never accurate enough for initial conditions for precision cosmology, especially if one is interested in higher-order statistics such as the matter bispectrum.
	\item Lower-order LPT with much higher starting redshift, e.g. $z_{\rm start}=199$, leads to much worse agreement with the reference solution. Instead, it indicates a ``spurious'' convergence to the discrete (lattice) solution, as opposed to the continuous fluid solution. It is also possible that early starts are further impacted by additional numerical force errors due to the smallness of perturbations.
	\item The impact of the truncation error increases with the order of the statistics considered: kurtosis is more affected than skewness, and bispectra are more affected than power spectra. One can conjecture that this continues to be even more severe for higher orders.
	\item On scales smaller than the quasi-linear scales, i.e., those which are dominated by halo profiles, we find a very weak dependence on initial conditions. As a consequence, the accuracy always improves at late times, while errors are largest at high redshift.
\end{enumerate}
Therefore, we deem it quite possible that the very high particle count per simulation volume size typically advocated for precision era simulations, is to some significant degree driven by discreteness errors due to early starts.
Specifically, to reach one per cent accurate predictions for the power spectrum at $k \sim 1\,\ihMpc$ over $z \in [0,1]$, it is required to employ LPT at 2nd order or higher with starting redshift $z \sim 50$ or less. At $k \sim 3\,\ihMpc$ the requirement is stronger, and only 3LPT is suitable for such accuracy. These numbers are however specific to the relatively low mass resolution that we employed in this paper, and at higher mass resolution one can expect altered requirements. We leave it for future work to establish the precise requirements for a given mass resolution.

In conclusion, our results thus suggest that late-start ICs based on higher-order LPT
might allow for more economic simulations to achieve a given accuracy -- a conjecture that should however be backed up with further convergence and resolution studies, and in particular also beyond 3LPT. Nevertheless, we expect that the increasing computational footprint at successively higher orders in LPT should eventually become too expensive compared to the improvement. We will come back to such issues in~\cite{RampfHahn2020}.

While our current results only concern simulations of the total matter flow in terms of a single collisionless fluid, the real Universe is composed of various distinct components, in particular baryons. Such multi-component simulations are however prone to more theoretical uncertainties. Simulations that model baryons with Lagrangian methods (SPH or moving mesh, when coupling to gravity using \(N\)-body techniques) suffer from particle discreteness errors which appear to be even more severe than in the single-fluid case \citep[cf.][]{Angulo:2013,Valkenburg:2017,Bird:2020}.  While not suffering from particle discreteness errors, Eulerian hydrodynamical simulations are impacted by advection errors and intrinsic smoothing due to the order of the finite volume method that also leave an imprint on the power spectrum \citep[][see their Fig.~23]{Hahn:2011}. Extrapolating the logic of the present paper, one way out of the situation for both Lagrangian and Eulerian hydrodynamics would be to initialize multi-fluid simulations also as late as possible to minimize discretization effects. This, however, requires the development of higher-order multi-fluid initial conditions to begin with, and will be accompanied by further hurdles that arise when incorporating baryonic physics. Such avenues will be assessed in \cite{RampfUhlemannHahn2020,HahnRampfUhlemann2020}.

%%%%%%%%%%%%%%%%%%%%%%%%%%%%%%%%%%%%%%%%%%%%%%%%%%
\section*{Acknowledgements}

We thank Bruno Marcos for useful discussions, and Romain Teyssier, Volker Springel, Simon White, and Jens Stuecker for comments on the manuscript.
M.M., O.H.,  and R.A. acknowledge funding from the European Research Council
(ERC) under the European Union's Horizon 2020 research and innovation programmes:
Grant agreement No. 679145 (COSMO-SIMS) for M.M. and O.H.; and for R.A., grant agreement No.\ 716151 (BACCO).
C.R.\ is a Marie Sk\l odowska-Curie Fellow and acknowledges funding from the People Programme (Marie Curie Actions)
of the European Union H2020 Programme under Grant Agreement No. 795707 (COSMO-BLOW-UP).
This work was granted access to the HPC resources of TGCC/CINES under the allocation A0060410847
attributed by GENCI (Grand Equipement National de Calcul Intensif). We thank the authors of {\sc BSkit} for making their software publicly available.

%%%%%%%%%%%%%%%%%%%%%%%%%%%%%%%%%%%%%%%%%%%%%%%%%%
%%%%%%%%%%%%%%%%%%%%%%%%%%%%%%%%%%%%%%%%%%%%%%%%%%
\section*{Data Availability}

The code to generate 3LPT initial conditions is freely  available at \url{https://bitbucket.org/ohahn/monofonic}. The data underlying this article will be shared on reasonable request to the corresponding author.

%%%%%%%%%%%%%%%%%%%% REFERENCES %%%%%%%%%%%%%%%%%%

\bibliographystyle{mnras}
\bibliography{bibliography}

%%%%%%%%%%%%%%%%% APPENDICES %%%%%%%%%%%%%%%%%%%%%

\appendix

\section{Explicit Formulae for 2LPT and 3LPT scalar and vector potentials}\label{app:LPT}
For convenience and reference, we give here the explicitly spelled out expressions for the $n$LPT potentials up to $n=3$:
\begin{align}
	\nabla^2 \Phi^{(2)} = & \,\Phi^{(1)}_{,xx}\left( \Phi^{(1)}_{,yy} + \Phi^{(1)}_{,zz}\right) + \Phi^{(1)}_{,yy}\Phi^{(1)}_{,zz}               \\
	                      & - \Phi^{(1)}_{,xy}\Phi^{(1)}_{,xy} - \Phi^{(1)}_{,xz}\Phi^{(1)}_{,xz}- \Phi^{(1)}_{,yz}\Phi^{(1)}_{,yz} \,,\nonumber
\end{align}
\begin{align}
	\nabla^2 \Phi^{(3a)} = & \,\Phi^{(1)}_{,xx}\Phi^{(1)}_{,yy}\Phi^{(1)}_{,zz} + 2 \Phi^{(1)}_{,xy}\Phi^{(1)}_{,xz}\Phi^{(1)}_{,yz}                                                             \\
	                       & - \left(\Phi^{(1)}_{,yz}\right)^2\Phi^{(1)}_{,xx} - \left(\Phi^{(1)}_{,xz}\right)^2\Phi^{(1)}_{,yy} - \left(\Phi^{(1)}_{,xy}\right)^2\Phi^{(1)}_{,zz} \,, \nonumber
\end{align}
\begin{align}
	\nabla^2 \Phi^{(3b)} = & \,\frac{1}{2}\Phi^{(1)}_{,xx}\left(\Phi^{(2)}_{,yy}+\Phi^{(2)}_{,zz}\right)                                                                                      \\
	                       & +\frac{1}{2} \Phi^{(1)}_{,yy}\left(\Phi^{(2)}_{,zz}+\Phi^{(2)}_{,xx}\right) + \frac{1}{2}\Phi^{(1)}_{,zz}\left(\Phi^{(2)}_{,xx}+\Phi^{(2)}_{,yy}\right)\nonumber \\
	                       & - \Phi^{(1)}_{,xy}\Phi^{(2)}_{,xy} -\Phi^{(1)}_{,xz}\Phi^{(2)}_{,xz}-\Phi^{(1)}_{,yz}\Phi^{(2)}_{,yz} \,, \nonumber
\end{align}
\begin{align}
	\nabla^2 A^{(3c)}_x = & \,\Phi^{(2)}_{,xy}\Phi^{(1)}_{,xz}-\Phi^{(2)}_{,xz}\Phi^{(1)}_{,xy} \\
	                      & +\Phi^{(1)}_{,yz}\left( \Phi^{(2)}_{,yy}-\Phi^{(2)}_{,zz}\right)
	-\Phi^{(2)}_{,yz}\left(\Phi^{(1)}_{,yy}-\Phi^{(1)}_{,zz} \right) \nonumber                  \\
	\nabla^2 A^{(3c)}_y = & \,\Phi^{(2)}_{,yz}\Phi^{(1)}_{,yx}-\Phi^{(2)}_{,yx}\Phi^{(1)}_{,yz} \\
	                      & +\Phi^{(1)}_{,zx}\left( \Phi^{(2)}_{,zz}-\Phi^{(2)}_{,xx}\right)
	-\Phi^{(2)}_{,zx}\left(\Phi^{(1)}_{,zz}-\Phi^{(1)}_{,xx} \right) \nonumber                  \\
	\nabla^2 A^{(3c)}_z = & \,\Phi^{(2)}_{,zx}\Phi^{(1)}_{,zy}-\Phi^{(2)}_{,zy}\Phi^{(1)}_{,zx} \\
	                      & +\Phi^{(1)}_{,xy}\left( \Phi^{(2)}_{,xx}-\Phi^{(2)}_{,yy}\right)
	-\Phi^{(2)}_{,xy}\left(\Phi^{(1)}_{,xx}-\Phi^{(1)}_{,yy} \right) \,. \nonumber
\end{align}
These expressions are equivalent to those in Eqs.\,\eqref{eq:lpt.potentials}.

%%%%%%%%%%%%%%%%%%%%%%%%%%%%%%%%%%%%%%%%

\section{Impact of Aliasing Errors on Late-Time Results}\label{appendix:aliasing}
\begin{figure}
	\centering
	\includegraphics[clip,type=pdf, ext=.pdf, read=.pdf, width=1.0\columnwidth]{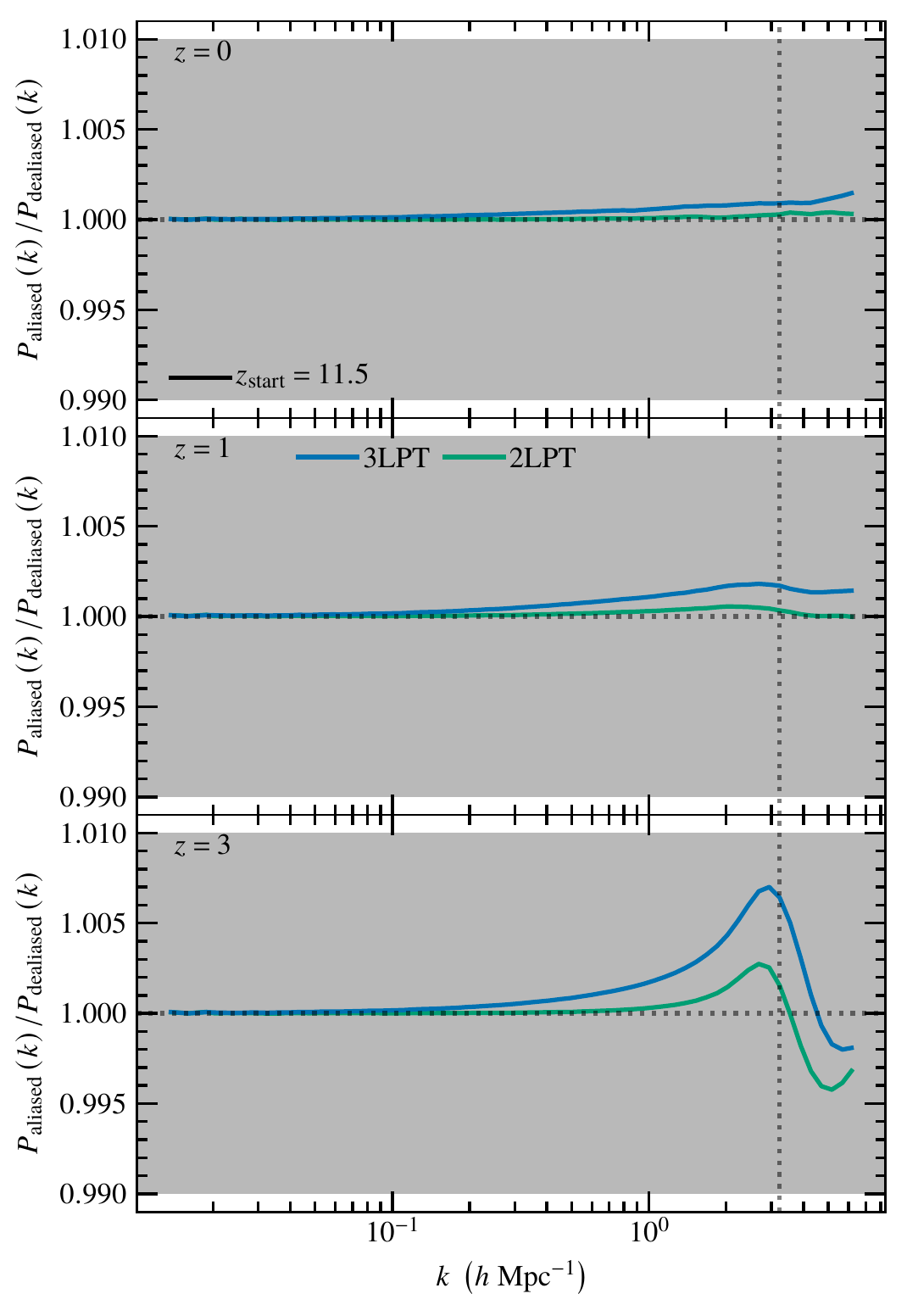}
	\caption{Comparison of low-redshift power spectra starting from initial conditions whose non-linearities have been properly de-aliased and those with aliasing errors for 2LPT and 3LPT starting at \(z_{\rm start}=11.5\) to maximise the error. Aliasing errors appear as decaying mode transients and thus, up to 3LPT, impact the power spectrum at the sub per-cent level at redshifts below 3. As before, the shaded grey area indicates a one per cent error.}
	\label{fig:powerspec.aliasing}
\end{figure}

In \autoref{fig:powerspec.aliasing}, we show the impact of aliasing errors in the non-linear 2LPT and 3LPT terms on the power spectrum at late times. Generally and as expected, aliasing errors become increasingly more relevant with higher order non-linearities, so the errors are larger in 3LPT than in 2LPT. Since they only appear in those higher order LPT terms, but not in the ZA, and since these terms are always subdominant on larger scales, we expect the impact to be weakened by non-linear evolution and not amplified. This is indeed what the simulations show: We find that at \(z=0\), the error is only \(\sim0.1\) per cent in the power spectrum, while at \(z=3\) the error is just under one percent, sharply peaking at the Nyquist mode. For the bispectrum we arrive at fairly similar conclusions (not shown).
It is thus most likely safe to work with an aliased version of 2LPT and 3LPT as long as one uses it as input for a fully non-linear simulation. Instead, if the output of nLPT were directly used, then the aliasing errors can be more significant, especially close to shell crossing.

%%%%%%%%%%%%%%%%%%%%%%%%%%%%%%%%%%%%
\section{Discreteness Effects and Lattice Evolution}\label{sec:plt}

Following \cite{Joyce:2005,Joyce:2007,Garrison:2016}, we consider the `particle linear theory' (PLT) of a perturbed simple cubic (SC) lattice.
Since our methodology for correcting the spurious PLT excitations differs slightly from \cite{Garrison:2016}, we begin with a review of the PLT approach which allows us to discuss the subtleties of the numerical implementations.

For simplicity, let us assume an EdS universe and consider a simple cubic lattice of $N$ particles, with identical mass, in a box with periodic boundary conditions.  The resulting equation of motion can be written as
\begin{align}
	\begin{aligned}
		 & \partial_a^2 \vect{x}_i = - \frac{3}{2a} \left(  \partial_a \vect{x}_i +  \nabla_x \varphi_N \right) \,,
	\end{aligned}
\end{align}
with $\vect{x}_i$ being the current comoving position of particle $i$ with initial position~$\vect{q}_i$, and $\varphi_N$ is the {\it  peculiar} gravitational potential, i.e., involving the usual subtraction of the overall background density; see e.g.\ Section 4 in \cite{Marcos:2006cn} for details. In PLT we are primarily interested to investigate the linear effects that stem from the particles discretization of the Poisson source. Expanding the Poisson source to linear order one gets
\begin{equation}
	\nabla_x \varphi_N(\vect{x}(\vect{q}_i)) =  - \frac 1 a \sum_{\vect{q}_j} {\cal D} (\vect{q}_i - \vect{q}_j) \, \vect{\psi}^{(1)}(\vect{q}_j,a) + {\cal O}({\psi^{(1)}}^2) \,,
\end{equation}
where ${\cal D}$ is the so-called dynamical matrix known in solid state physics (cf.\ \citealt{Marcos:2006cn} and references therein).

In Fourier space, the evolution equation for the linear displacement can then be written as (suppressing the particle index)
\begin{equation}
	a^2 \partial_a^2 \tilde{\vect{\psi}}^{(1)}(\vect{k})  + (3a/2) \,\partial_a \tilde  {\vect{\psi}}^{(1)}(\vect{k})  - (3/2) \, \tilde {\cal D}(\vect{k})\,\tilde  {\vect{\psi}}^{(1)}(\vect{k}) = 0 \,.
\end{equation}
To make progress, we follow \cite{Marcos:2006cn}
and solve the above equation  with the {\it Ansatz}
\begin{align}\label{ansatzPLT}
	\tilde  {\vect{\psi}}^{(1)}(\vect{k},a) = \sum_{n=1}^3  \hat{\mathbf{e}}_n(\vect{k}) \, \tilde \psi_n(\vect{k},a) \,,
\end{align}
leading to  an eigenvalue problem as well as an ODE for  $\tilde  \psi_n$, which respectively read
\begin{align}\label{ODE}
	\begin{aligned}
		 & \tilde {\cal D}(\vect{k}) \,\hat{\mathbf{e}}_n(\vect{k}) = \varepsilon_n(\vect{k})\, \hat{\mathbf{e}}_n(\vect{k}) \,,                 \\
		 & a^2 \partial_a^2 \tilde  \psi_n   +(3a/2) \, \partial_a  \tilde  \psi_n  - (3/2) \, \varepsilon_n(\vect{k}) \,\tilde  \psi_n  = 0 \,,
	\end{aligned}
\end{align}
where the eigenvalues $\varepsilon_n$ can be determined using standard linear algebra programs (see further below for details).
The ODE has the general solution
\begin{equation}
	\tilde \psi_n = C^+_n a^{\alpha^+_n(\vect{k})} + C^-_n a^{\alpha^-_n(\vect{k})} \,,
\end{equation}
where $C^{+/-}_n$ are integration constants, and $\alpha_n^\pm(\vect{k}) = (1/4) [-1 \pm \sqrt{1+ 24 \varepsilon_n(\vect{k})}]$.
In the following two paragraphs we shall enforce that the PLT solution is compatible with purely growing mode  and curl-free solutions. After that, we outline some technical details about the numerical computation of the dynamical matrix.

\paragraph*{Demanding curl-free solutions.} One of the eigenmodes called $\hat{\mathbf{e}}_\parallel$ is exactly parallel to $\hat{\vect{k}}$ in the fluid limit, and thus, $\hat{\mathbf{e}}_\parallel$ is associated with gravity induced modes. The other two eigenmodes are orthonormal to $\hat{\mathbf{e}}_\parallel$, and, in the fluid limit, are associated with transverse modes that are vectorial transients and fictitious in the fluid sense, due to the assumed irrotational character of CDM. Thus, to initialize \(N\)-body simulations, we need to make sure that only the lattice modes along $\hat{\mathbf{e}}_\parallel$ are excited; to achieve this, we rescale the $\vect{k}$ vectors appearing in the LPT fluid solution $\tilde \psi_{\rm fluid} = {\rm i} \vect{k} \tilde \phi + {\rm i} \vect{k} \times \tilde{\vect{A}}$  according to \cite[cf.][]{Garrison:2016}
\begin{equation}
	\vect{k} \to \hat{\mathbf{e}}_\parallel /(\hat{\mathbf{e}}_\parallel \cdot \hat {\vect{k}})\,,
\end{equation}
where the division of the scalar product is included to ensure that the matter power spectrum has unchanged amplitudes.
This procedure, applied to ZA, has been first outlined by \cite{Garrison:2016}, although we note that we apply this rescaling  also beyond ZA.

\paragraph*{Demanding purely growing-mode solutions.} In the fluid limit, it can be shown that $\alpha^+_\parallel \to 1$ and $\alpha^-_\parallel \to -3/2$ and thus, the solution of the ODE that is proportional to $C_-$ is a decaying one, and thereby not compatible with growing-mode solutions.
To enforce growing-mode solutions, we first evaluate $\tilde \psi_\parallel = C^+_\parallel a^{\alpha^+_\parallel} + C^-_\parallel a^{\alpha^-_\parallel}$ and its first time derivative at initial time $a_{\rm ini}$, and express the integration constants $C^{+/-}_\parallel $  in terms of $\tilde \psi_\parallel(a_{\rm ini}) \equiv \tilde \psi_{\rm ini}$ and $\dot{\tilde \psi}_\parallel(a_{\rm ini}) \equiv \tilde v_{\rm ini}$. We find that
\begin{equation}
	C^-_\parallel  \propto  \frac{\tilde v_{\rm ini}}{\alpha_\parallel^+} -  \frac{{\tilde \psi}_{\rm ini}}{a_{\rm ini}} \,,
\end{equation}
and thus, decaying modes in PLT are zero provided that the initial PLT velocity and displacement satisfy  ${\tilde v_{\rm ini}}/{\alpha_\parallel^+} =  {\tilde \psi}_{\rm ini}/{a_{\rm ini}}$. This relation implies that only a single initial potential (e.g., $v_{\rm ini}$, ${\psi}_{\rm ini}$, or the initial gravitational potential) needs to be specified; this is precisely expected for slaved initial conditions.
Comparing this with the fluid case for which the slaved initial conditions imply
\begin{equation}
	C^-_{\rm fluid} \propto \tilde v_{\rm ini}^{\rm fluid} - \frac{{\tilde \psi}_{\rm ini}^{\rm fluid}}{a_{\rm ini}} \to 0 \,,
\end{equation}
where $\vect{v}^{\rm fluid} =\nabla v^{\rm fluid}$ and $\vect{\psi}^{\rm fluid} =\nabla \psi^{\rm fluid}$,
we realize that growing-mode PLT initial solutions in the \(N\)-body simulation are established provided we rescale the fluid velocity according to
\begin{equation}
	\tilde{\vect{v}}_{\rm ini}^{\rm fluid} \to  \tilde{\vect{v}}_{\rm ini}^{\rm fluid}/\alpha_\parallel^+ \,.
\end{equation}
We found that this somewhat ad-hoc modification, demonstrated to be strictly correct at leading order, works very well also in combination with higher-order LPT, as demonstrated by our results shown in \autoref{fig:powerspec.plt.sc.convergence} and \autoref{fig:bispec.equilateral.plt}.

\paragraph*{Numerical evaluation of dynamical matrix.}
Following \cite{Marcos:2006cn}, we compute the dynamical matrix $\mathcal{D}_{ij}$ using Ewald summation, i.e.\ we split $\mathcal{D}$ into a short-range and a long-range contribution, $\mathcal{D}_{ij} = \mathcal{D}^{\textrm{sr}}_{ij}+\mathcal{D}^{\textrm{lr}}_{ij}$ with splitting parameter $\kappa$. Specifically for an SC lattice, we compute on a grid of coordinates in real space, with $r:=\left\| \vect{x}\right\|$
\begin{align}
	\mathcal{D}^{\textrm{sr}}_{ij}(\vect{x}\neq \vect{0}) & =  -\frac{\kappa^3}{\uppi^{3/2}}\frac{x_i\,x_j}{r^2}{\rm e}^{-\kappa^2r^2}                                                                                                                    \\
	                                                      & + \frac{1}{4\uppi}\left( \frac{\delta_{ij}}{r^3} - 3\frac{x_i\,x_j}{r^5}\right) \left[\textrm{erfc}\left(\kappa r\right)+\frac{2\kappa}{\sqrt{\uppi}}{\rm e}^{-\kappa^2 r^2}\right] \nonumber
\end{align}
and we set $\mathcal{D}^{\textrm{sr}}_{ij}(\vect{x}= \vect{0}) = 0$. We then Fourier-transform this short-range response function using the FFT. The respective long-range component is given in Fourier space as
\begin{align}
	\widetilde{\mathcal{D}}^{\textrm{lr}}_{ij}(\vect{k}\neq\vect{0}) & =  \frac{ \delta_{ij}}{3} +  \frac{k_i k_j}{\left\| \vect{k} \right\|^2} {\rm e}^{ -\left\|\vect{k}\right\|^2/(4\kappa^2)},\quad\widetilde{\mathcal{D}}^{\textrm{lr}}_{ij}(\vect{k}=\vect{0}) =\frac{ \delta_{ij}}{3}  .
\end{align}
We use a cut-off scale corresponding to 4 cells, i.e.\ $\kappa=(\sqrt{2}\,4 \Delta x)^{-1}$. We sum short and long-range components to obtain the full dynamical matrix $\mathcal{D}_{ij}(\vect{k})$ in Fourier space, then diagonalize it to obtain the lattice growing mode eigenvectors $\vect{e}_\parallel$ with associated eigenvalue $\epsilon_\parallel$. Since the dynamical matrix is very smooth, and its computation for the actual full-resolution lattice relatively costly, we compute a relatively low resolution version of it (in fact $64^3$) and then interpolate to a higher resolution lattice (as also \citealt{Garrison:2016} are suggesting).

%%%%%%%%%%%%%%%%%%%%%%%%%%%%%%%%%%%%
\section{Invariance conservation as convergence and correctness test}\label{sec:invariance_test}

\begin{figure}
	\centering
	\includegraphics[type=pdf, ext=.pdf, read=.pdf, width=1.0\columnwidth]{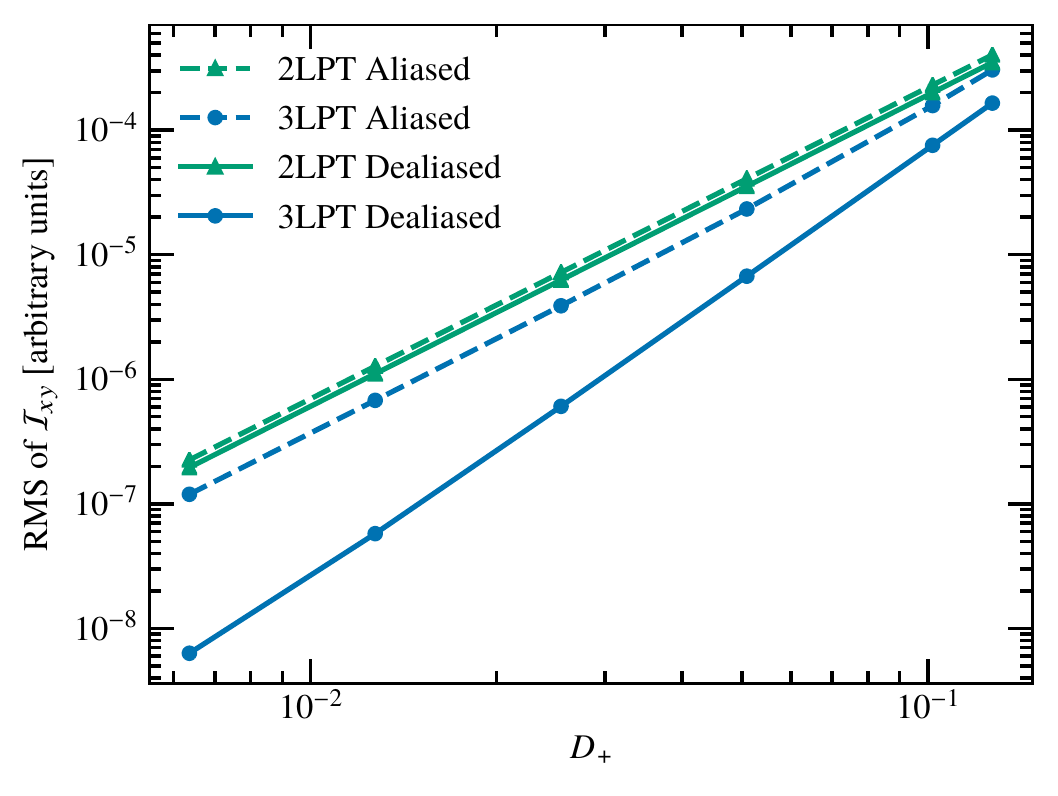}
	\caption{RMS amplitude of the violation of the Cauchy invariants~\eqref{eq:cauchy_invariants} as a function of starting time \(D_+\) for 2LPT and 3LPT. We plot the standard deviation of the \(\mathcal{I}_{xy}\) field for a cosmological \(1\,h^{-1}{\rm Gpc}\) box with \(1024^3\) modes and particles. The amplitudes for \(\mathcal{I}_{yz}\) and \(\mathcal{I}_{xz}\) are identical. We generated the LPT fields using both Orszag's 3/2 rule for de-aliasing (solid lines), and also not applying any de-aliasing procedure (dashed lines). Clearly only the de-aliased results scale \(\mathcal{O}( D_{+}^3 )\) for 3LPT, in all other cases we find behaviour \(\mathcal{O}( D_{+}^2 )\) that is expected for 2LPT. Note that \(\mathcal{I}_{ij}\) vanishes exactly for the Zel'dovich approximation. }
	\label{fig:invariants_evol}
\end{figure}

Testing numerical implementations of LPT in a realistic cosmological setup (i.e.\ with a wide spectrum of modes) is inherently difficult. Here we propose a new approach: we know that LPT violates certain symmetries of the fluid equations perturbatively beyond the ZA.
Specifically, \cite{Uhlemann:2019} have shown that beyond ZA, fixed-order LPT at order \(n\) excites artificial vortical modes at order \(n+1\). The relevant equations for this are the Cauchy invariants, which are the local form of Kelvin's circulation theorem.
In the cosmological context the Cauchy invariants state the conservation of the zero vorticity, which is satisfied at all times for individual fluid particles. The three invariants are (\(p=1,2,3\))
\begin{equation}
	C_p = \varepsilon_{pln}  v_{k,l} x_{k,n} = 0.
\end{equation}
These vectorial relations can be written in tensor form (thereby making contact with similar theorems in Hamiltonian mechanics).
Contracting $C_p$ with the Levi-Civita symbol $\varepsilon_{pij}$ yields the antisymmetric tensor
\begin{align}
	\mathcal{I}_{ij} := \varepsilon_{pij}   C_p & =  v_{k,i}\,x_{k,j} - v_{k,j}\,x_{k,i} \nonumber                                                 \\
	                                            & = v_{j,i}-v_{i,j} + v_{k,i}\,\psi_{k,j} - v_{k,j}\,\psi_{k,i} = 0\,,\label{eq:cauchy_invariants}
\end{align}
where we used that $x_{k,j} = \delta_{kj} + \psi_{k,j}$. Equation~\eqref{eq:cauchy_invariants} vanishes non-perturbatively. However, only in the case of the Zel'dovich approximation,  $\mathcal{I}_{ij}$ is exactly zero, when $v_i$ and $\psi_i$ are truncated at the perturbative order.
As mentioned before, for $n$th order LPT with $n>1$, the truncation error is of order $n+1$ \citep[cf.][]{Uhlemann:2019}. Since the Cauchy invariants contain a time derivative through $v_k = \dot x_k$, however, the resulting truncation error grows with $D_+^n$ \ (and not as $D_+^{n+1}$).
We show the RMS amplitude of $\mathcal{I}_{xy}$ for our implementation as a function of the starting time $D_+$ in Eq.\,\eqref{fig:invariants_evol}. We see that only when de-aliasing the non-linearities, we get the expected scaling of $\mathcal{I}_{ij}$ with $D_+^3$ in 3LPT demonstrating the correctness of our implementation.

On a technical note, numerically, in both cases, we applied 3/2 padding when evaluating Eq.\,\eqref{eq:cauchy_invariants}, but compute $v_i$ and $\psi_i$ with and without padding for second and higher order non-linearities to test the effect of aliasing on velocities and displacements.

Note that the lines for 2LPT and 3LPT in \autoref{fig:invariants_evol} intersect close to $a\sim0.1$ which is further evidence that this is the point where LPT breaks down, consistent with our results in Sec.~\ref{sec:convergence_radius}.

%%%%%%%%%%%%%%%%%%%%%%%%%%%%%%%%%%%%
\section{Details to obtaining the theoretical bound on the radius of convergence}\label{app:theoryconvergence}

Here we outline the derivation of the bounds on the theoretical radius of convergence of LPT, section~\ref{sec:theorybound}.

\subsection{Most conservative estimate}\label{app:veryconservative}

Details in this subsection are based on~\cite{Zheligovsky:2014,Rampf:2015}.
Beyond first order, the LPT recursion relations express the (to be determined) $n$th-order displacement gradients coefficients $\psi_{i,j}^{(n)}(\vect{q})$ in terms of quadratic and cubic combinations of (known) displacement gradients from lower orders.
These quadratic and cubic combinations of displacements contain numerical coefficients that are all bounded by unity from below. Furthermore, these quadratic and cubic displacement terms contain operators of the kind $\nabla^{-2} \partial_i \partial_j$; applying suitable norms on the recursion relation of the displacement gradients, i.e., by considering the recursion relation for $\| \psi_{i,j}^{(n)} \|$, it can be shown that also such operators are bounded by unity.
Exploiting those two bounds by setting the coefficients and operators effectively to unity, the recursion relation for the coefficient $\| \psi_{i,j}^{(n)} \|$  turns into a cubic {\it polynomial inequality}.  Upon reintroducing the time variable $D_+$ by using the generating function $\Psi \equiv \sum_n \| \psi_{i,j}^{(n)} \| \,D_+^n$,
this polynomial inequality turns into\footnote{For simplicity we assume here an Einstein--de Sitter universe which is a sufficient approximation for our purpose. \cite{Rampf:2015} generalized the presented results to the \(\Lambda\)CDM universe, however found only a weak dependence of \(\Lambda\) on the lower bound of convergence.}
\begin{equation}\label{poly}
	p_{\rm low}(D_+, \Psi) :=  6 \Psi^3 + 12 \Psi^2 - \Psi + D_+ \| \nabla_i \nabla_j \varphi_{\rm ini} \| \geq 0   \,,
\end{equation}
where $\varphi_{\rm ini}$ is given by Eq.\,\eqref{varphiini}.
For small $\Psi$'s, the polynomial $p_{\rm low}(0,\Psi)$ behaves as $-\Psi$ while asymptotically it behaves like $\Psi^3$.
Furthermore, for $D_+ > 0$,  the cubic polynomial is shifted upwards linearly in $D_+$ along the y-axis, and has first three roots and later on a single root.
To investigate the physical branch, let us begin at $D_+ =0$ for which there is a root crossing the origin, in accordance with a zero displacement initially and thus this root denotes the physical branch. For times shortly after, this root moves from zero to a small positive value, and furthermore the whole branch between 0 and that root is positive, thereby satisfying the positivity condition of~\eqref{poly} and thus marking the physical branch. At some critical time later, denoted $D_+^{\rm crit}$, this root will merge with another one. Just shortly after  $D_+^{\rm crit}$, these two roots disappear (i.e., by becoming complex), implying that $p$ loses its boundedness and so does the displacement field\footnote{The unboundness of the polynomial $p_{\rm low}$ essentially implies that the higher-order displacement coefficients do not decrease at subsequent orders -- which is a clear indication that the series lost its convergence property.}. Thus, $D_+^{\rm crit}$ marks the last possible time for which the displacement is bounded, and thus   is the maximal time for which convergence is still guaranteed, $D_+^{\rm crit} = D_+^{\rm theory}$. That maximal time can be determined by setting the discriminant of $p_{\rm low}(\Psi)$ to zero. This leads to the positive critical time
\begin{equation}
	\label{eq:dplus.theory.app}
	D_+^{\rm theory} = \frac{\rm T}{\| \nabla_i \nabla_j \varphi_{\rm ini} \|} \,,
\end{equation}
with ${\rm T} = 0.0204$ which, as promised, is a lower bound on the radius of convergence \mbox{$D_+^{\rm theory} < R$.} In the following we show how this bound can be improved.

\subsection{Conservative, third-order estimate}\label{app:third}

As outlined above, before obtaining the polynomial inequality~\eqref{poly} from which the lower bound on the convergence is retrieved, we set all numerical coefficients in front of the nth-order displacement to unity, simply since those coefficients are bounded by unity. However, we do have explicit numerical coefficients up to 3LPT (cf.\ Eq.\,\ref{eq:3lpt_solutions}). A refinement for the theoretical bound is therefore obtained by keeping some of these coefficients as they really are, and only ``set'' the higher-order coefficients to unity. Doing so leads after some straightforward calculations to the modified polynomial inequality
\begin{equation}\label{poly2}
	p_{\rm 3LPT}(D_+, \Psi) :=  6 \Psi^3 + 12 \Psi^2 - \Psi + {\rm T} - \frac{24} 7 {\rm T}^2 -\frac{86}{7} {\rm T}^3  \geq 0   \,,
\end{equation}
where ${\rm T} := D_+ \| \nabla_i \nabla_j \varphi_{\rm ini} \|$, which, by identification of the result from the previous section leads to the equation
${\rm T} - 24{\rm T}^2/7 -86 {\rm T}^3/7 = 0.0204$, leading to
\begin{equation}
	{\rm T} =:   {\rm T}_{\rm cons} = 0.022
\end{equation}
for Eq.\,\eqref{eq:dplus.theory.app}, a $9$\,\% improvement regarding the previous section.

\subsection{Optimistic (shell-crossing) estimate}\label{app:optmistic}

Observe that shell-crossing occurs when the Jacobian $J = \det[\delta_{ij}+\psi_{i,j}] = 1 + \nabla \cdot \psi + (1/2)[\psi_{i,i} \psi_{j,j} - \psi_{i,j}\psi_{i,j}] + \det[\psi_{i,j}]$
vanishes for the first time. As it was shown by~\cite{Rampf:2015}, by the same methods as explained in appendix~\ref{app:veryconservative}, we can obtain a bound on the vanishing Jacobian by introducing the generating function~$\Psi =\sum_n \| \psi_{i,j}^{(n)} \| \,D_+^n$. One finds
\begin{equation}\label{Jbound}
	|J-1| \leq 6 \Psi^3 + 6 \Psi^2 + 3\Psi \equiv Q(\Psi) \,.
\end{equation}
Now, from the bound as obtained  in appendix~\ref{app:veryconservative}, one can easily determine the corresponding critical value $\Psi_{\rm crit} \simeq 0.0404$ for which the bound~\eqref{eq:dplus.theory.app} holds. Plugging this value in Eq.\,\eqref{Jbound}, one finds $|J-1| \leq 0.132$, which is well below unity for which shell-crossing occurs. Thus, for the  conservative bound $D_+^{\rm theory}$, shell-crossing is not reached.
Therefore, we seek here in this section to obtain stronger bounds, thereby allowing us to reach shell-crossing closely from below.

In the main text, we have mentioned that the first singularity in LPT could be in the complex time-domain; however, recently~\cite{Saga:2018} provided solid numerical evidence that LPT converges until shell-crossing. Specifically, they estimated the radius of convergence for sine-wave initial conditions in 3D, by using LPT to 10th order. Comparison to Vlasov--Poisson simulations revealed indeed that LPT appears to converge until shell-crossing, excluding the spherical (or symmetric sine-wave) collapse which is degenerate in a universe with random initial conditions.

Thus, in our language, a complex singularity is unlikely to be decisive for determining the radius of convergence for sine-wave ICs, and the same should also be true for generic, Gaussian initial conditions.

Therefore, in this appendix we {\it assume} that complex singularities
are ruled out, and that the convergence-determining factor is indeed
shell-crossing. One reason why the methods in
appendix~\ref{app:veryconservative} did not allow us to come closer to
the first shell-crossing, is the failure to provide better estimates
for the all-order coefficients and spatial derivatives, that in
particular sensitively influence some prefactors within the polynomial
inequality, Eq.\,\eqref{poly}. Specifically, the following altered polynomial
inequality -- which is not based on a theory but rather on intuition
-- would be in principle closer to the true inequality for the
generating function,
\begin{equation}
	p_{\rm cross}(D_+, \Psi) := 6 a \Psi^3 + 12 b \Psi^2 - \Psi + {\rm T}_{\rm cross} \geq 0 \,, \end{equation}
where $a, b$ are positive but a priori unknown coefficients, whereas
${\rm T}_{\rm cross}$ is the critical coefficient to be determined, since it sets
the (possibly strongest) bound on the time of shell-crossing.

Now, we first search for the maximal value of $\Psi_{\rm max}$ for which the bound on the Jacobian just
begins to touch the unity bound, i.e., $Q(\Psi_{\rm max}) = 1$. We find $\Psi_{\rm max} = 0.2178$.
If the $p_{\rm cross}$ has its critical value $\Psi_{\rm crit}$ for which its discriminant vanishes at precisely
$\Psi_{\rm max}$, then ${\rm T}_{\rm cross}$ would be maximized.
Thus,  solving for $p_{\rm cross}(\Psi_{\rm max}) =0 = p_{\rm cross}'(\Psi_{\rm max})$ leads to updated constraints.
Unfortunately, the system of equations is still under determined -- in fact there is a redundant free parameter (e.g., either $a$ or $b$),
thereby seemingly rendering the considered task as unsolvable.

Instead of solving explicitly for the two coefficients $a$ and $b$, however, we could simply scan through the allowed parameter space where
shell-crossing is ruled out, as well the two constrains $a>0$ and $b>0$ are satisfied. Scanning through the parameter space,
we arrive at the (optimistic) theoretical shell-crossing constraint
\begin{equation}
	D_+^{\rm cross} = \frac{{\rm T}_{\rm cross}}{\| \nabla_i \nabla_j \varphi_{\rm ini} \|} \,,
\end{equation}
with the worst-case coefficient in the allowed regime to be ${\rm T}_{\rm cross} = 0.107$.

%%%%%%%%%%%%%%%%%%%%%%%%%%%%%%%%%%%%%%%%%%%%%%%%%%

% Don't change these lines
\bsp	% typesetting comment
\label{lastpage}
\end{document}